\documentclass[preprintnumbers,showpacs,showkeys,preprint,secnumarabic,amssymb,amsmath,amsfonts,aps, pre]{revtex4-1}

\usepackage{graphicx}

\usepackage{amsmath}
\usepackage{amsfonts}
\usepackage{amssymb}

\newcommand{\beq}{\begin{equation}}
\newcommand{\eeq}{\end{equation}}
\newcommand{\bea}{\begin{eqnarray}}
\newcommand{\eea}{\end{eqnarray}}
\newcommand{\bml}{\begin{mathletters}}
\newcommand{\eml}{\end{mathletters}}

\newcommand{\f}{\begin{equation}}
\newcommand{\ff}{\end{equation}}
\newcommand{\be}{\begin{equation}}
\newcommand{\ee}{\end{equation}}
\newcommand{\ba}{\begin{array}}
\newcommand{\ea}{\end{array}}
\newcommand{\bena}{\begin{eqnarray}}
\newcommand{\eena}{\end{eqnarray}}
\newcommand{\bdis}{\begin{displaymath}}
\newcommand{\edis}{\end{displaymath}}
\newcommand{\bit}{\begin{itemize}}
\newcommand{\eit}{\end{itemize}}
\newcommand{\ben}{\begin{enumerate}}
\newcommand{\een}{\end{enumerate}}
\newcommand{\virgolette}[1]{``#1''}

\newcommand{\nero}[1]{\boldsymbol{#1}}
\newcommand{\pieno}[1]{\boldsymbol{#1}}

\everymath{\displaystyle}

\begin{document}
%%%%%%%%%%%%%%%%%%%%%%%%%%%%%%%%%%%%%%%%%%%%%%%%%%%%%%%%%
%Eliminating nonphysical instabilities of geodesic flows of Jacobi metric
% AUTHOR INFORMATIONS
\title{Coherent Riemannian-geometric description of Hamiltonian order and chaos with Jacobi metric}

\date{\today}

\author{Loris Di Cairano}
\email{l.di.cairano@fz-juelich.de}
\affiliation{Computational Biomedicine, Institute for Advanced Simulation IAS-5, and Institute of Neuroscience and Medicine INM-9, Forschungszentrum Jülich, 52425 Jülich, Germany}
\affiliation{Center for Computational Engineering Science, Department of Mathematics, RWTH Aachen University, Germany}

\author{Matteo Gori}
\email{gori6matteo@gmail.com}
\affiliation{Aix-Marseille University, CNRS Centre de Physique Th\'eorique UMR 7332,
Campus de Luminy, Case 907, 13288 Marseille Cedex 09, France}

\author{Marco Pettini}
\email{pettini@cpt.univ-mrs.fr}
\affiliation{Aix-Marseille University, CNRS Centre de Physique Th\'eorique UMR 7332,
Campus de Luminy, Case 907, 13288 Marseille Cedex 09, France}

%%%%%%%%%%%%%%%%%%%%%%%%%%%%%%%%%%%%%%%%%%%%%%%%%%%%%%%%
% ABSTRACT AND PACS
\begin{abstract}
By identifying Hamiltonian flows with geodesic flows of suitably chosen Riemannian manifolds, it is possible to explain the origin of chaos in classical Newtonian dynamics and to quantify its strength. There are several possibilities to geometrize Newtonian dynamics under the action of conservative potentials and the hitherto investigated ones provide consistent results. However, it has been recently argued that endowing configuration space with the Jacobi metric is  inappropriate to consistently describe the stability/instability properties of Newtonian dynamics because of the non-affine parametrization of the arc length with physical time. To the contrary, in the present paper it is shown that there is no such inconsistency and that the observed instabilities in the case of integrable systems using the Jacobi metric are artefacts.
\end{abstract}
%%%%%%%%%%%%%%%%%%%%%%%%%%%%%%%%%%%%%%%%%%%%%%%%%%%%%%%%%
\maketitle
%%%%%%%%%%%%%%%%%%%%%%%%%%%%%%%%%%%%%%%%%%%%%%%%%%%%%%%%%
% BODY OF PAPER
%%%%%%%%%%%%%%%%%%%%%%%%%%%%%%%%%%%%%%%%%%%%%%%%%%%%%%%%%
\section{Introduction} 
Elementary tools of Riemannian differential geometry can be successfully used to explain the origin of chaos in Hamiltonian flows or, equivalently, in Newtonian dynamics.
Natural motions of Hamiltonian systems can be viewed as geodesics of the configuration-space manifold $M$ equipped with the Riemannian metric $g_J$, known as the Jacobi metric (or kinetic energy metric). The stability/instability properties of such geodesics can be investigated by means of the Jacobi--Levi-Civita (JLC) equation for geodesic spread. It has been shown that chaos in physical geodesic flows does not stem from hyperbolicity of $M$:  phase space trajectories/geodesics  are destabilized both by regions of negative curvature and by parametric instability caused by positive curvature varying along the geodesics \cite{marco,physrep,book}. Another remarkable fact is that the JLC equation written for a geometrization of Hamiltonian systems in an enlarged configuration space-time endowed with a metric due to Eisenhart  \cite{eisenhart} yields the standard tangent dynamics equation commonly used in numerical computations of the largest Lyapunov exponent (LLE). These two different geometric framework have been proven to give the same information about order and chaos and about the strength of chaos as well. This has been
checked in the case of two-degrees of freedom Hamiltonian systems, for the H\'enon-Heiles model and for two coupled quartic oscillators, respectively   \cite{cerruti1996geometric,rick},
and  in the case of a large number of degrees of freedom (from 150 up to 1000) \cite{cerruti1997lyapunov}. It has been found that the JLC equation stemming from the Jacobi metric gives exactly the same  quantitative results of the tangent dynamics equation.
 
This notwithstanding, in Ref. \cite{cuervo2015non} it has been argued that the non-affine parametrization of the arc-length with time in configuration space endowed with the Jacobi metric  leads to nonphysical instabilities. More precisely,  the JLC equation written for the Jacobi metric seems to give chaos also for a system of harmonic oscillators. Moreover, such alleged non-physical instabilities are found to be stronger in systems with few degrees of freedom.
Such an argument seems to radically exclude the use of Jacobi metric in configuration space to consistently investigate Hamiltonian chaos; in fact, the non-affine parametrization of the arc length with respect to physical time is an unavoidable consequence of the way Jacobi metric is derived  from Maupertuis' least action principle.
Some mathematical works \cite{montgomery2014s,giambo2014morse,giambo2015normal}, partly motivated by the results reported in Ref.\cite{cuervo2015non}, have investigated the behaviour of the geodesics in configuration space endowed with the Jacobi metric near the so called Hill's boundaries, i.e. the regions in configuration space where $E=V(\mathbf{q})$ and where the Jacobi metric is singular ($g_{J}=0$). In particular, all these works emphasized the phenomenon of geodesic reflection near Hill's boundaries and the relevance of  these reflections to characterize periodic orbits, with, among the others,  an original proposal dating back to Ref. \cite{seifert1948periodische}.
Finally, other authors \cite{yamaguchi2001geometric} have tackled  the geometrization of Hamiltonian dynamical systems by lifting the Jacobi metric from configuration space $M$ to its cotangent bundle $T^{*}M$, i.e. to the phase space. In this framework JLC equations are rewritten as a system of first order linear differential equations on the tangent bundle $TT^{*}M$ of phase space: this allows to identify the adequate degrees of freedom to compute the \emph{Geometric Largest Lyapunov Exponent}(GLLE). Within this framework, the GLLE (JLC equations) and LLE (tangent dynamics) are found in very good agreement to characterize the chaotic regime (strong/weak chaos) of the H\'enon-Heiles model.
All these studies, together with the manifest contradiction among the outcomes of Ref.\cite{cuervo2015non} and those of Refs. \cite{cerruti1996geometric,rick} have motivated the present work.    
The present paper is organized as follows.
In Section \ref{sec:confscal_jacobi} we briefly discuss some aspects of the construction of the Jacobi metric, with emphasis on the consequences of non-affine parametrization of the arc length with time for the JLC equation, and on the presence of boundaries where the metric is singular. 
In Section \ref{sec_parallel_transported_frame}, the JLC equation is rewritten by introducing a parallel transported frame for which explicit expressions are then given  in Subsections  \ref{subsec_simulation_parallel_transported_frame2} and  \ref{subsec_simulation_parallel_transported_frame3} for $N=2$ and $N=3$, respectively. Then in Subsections \ref{simulation2} and \ref{simulation3} the results of the corresponding numerical simulations are reported for two and three (resonant) harmonic oscillators, respectively, using the most critical values of the parameters which, according to Ref.\cite{cuervo2015non}, should yield non physical instabilities. It is shown that this is not the case.
In Section \ref{sec_concentration_measure}, we discuss how the measure concentration phenomenon, which takes place at a large number of degrees of freedom, completely removes any problem making unnecessary even resorting to a parallel transported frame.
Some conclusions are drawn in Section \ref{conclusion}. 
\section{Effects of Non-Affine Parametrization of the arc length with Jacobi Metric}
\label{sec:confscal_jacobi}
Among the different possibilities of rephrasing Newtonian dynamics in geometric terms, as reported in \cite{physrep,book}, 
the Jacobi metric in configuration space leads to the mathematically richest structure [$(M,g_J)$ is geodesically complete in the sense of the Hopf-Rinow theorem]. 
Let us consider a thorough investigation of the geometrization of Newtonian dynamics by means of $(M,g_J)$ for systems described by a Lagrangian of the form
\begin{equation}
L(\dot{q},q)=\dfrac{1}{2} \overline{g}_{ij}(q) \dot{q}^{i} \dot{q}^j -V(q)\,\,\, ,
\end{equation}
where
\begin{equation}
\dot{q}^{i}(t)=\dfrac{\mathrm{d}q^i}{\mathrm{d}t}(t)
\end{equation}
$\overline{g}_{ij}$ are the components of the kinetic energy metric on configuration space $M$ with the associated Levi-Civita connection $\overline{\nabla}$ specified by the Christoffel coefficients $\overline{\Gamma}^{i}_{jk}$, and $V(q)$ is the potential energy.
It is well known that, given a chart $({U},\phi)$ on the configuration space $M$ and a curve $\gamma:I\subset\mathbb{R}\longrightarrow M$, the natural motions $\phi(\gamma(t))=\pieno{\gamma}(t)=\pieno{q}(t)$ are the class of curves that make stationary the
action functional
\begin{equation}
\label{eq:action_functional}
S[\pieno{q}(t)]=\int_{t_0}^{t_1} L(\dot{\pieno{q}}(t),\pieno{q}(t)) \mathrm{d}t
\end{equation} 
on the class of curves with $\pieno{q}(t_0)=\pieno{a}$ and $\pieno{q}(t_1)=\pieno{b}$ fixed, i.e.
\begin{equation}
\label{eq:leastact_princ}
\delta S[\gamma]=0 \qquad \text{with} \qquad \gamma(t_0)=a \,\,\,\, \text{and}\,\,\,\,\, \gamma(t_1)=b \ .
\end{equation}
 The Newton equations are derived from the Euler-Lagrange equations, i.e.
\begin{equation}
\overline{\nabla}_{\dot{\gamma}}\dot{\gamma}=-\overline{\mathrm{grad}} V
\end{equation}
where $\overline{\mathrm{grad}}^{i} f(q)=\overline{g}^{ik}\partial_{k} f$ is the gradient.
Natural motions $\gamma(t)$ belong to a class of curves of configuration space satisfying  the "physical" variational principle \eqref{eq:leastact_princ},
therefore these curves can be identified with geodesics  of configuration space which also satisfy a variational principle but in this case of a  "geometrical" kind. In fact, geodesics  $\tilde{\gamma}(s)$ are curves of a Riemannian manifold $(M,\tilde{g})$ endowed with a metric $\tilde{g}$ that makes stationary the length functional between two fixed points, i.e.
\begin{equation}
\delta\mathit{l}[\tilde{\gamma}]=0 \qquad \text{with} \quad \mathit{l}[\tilde{\gamma}(s)]=\int_{s_{0}}^{s_{1}}\sqrt{\tilde{g}_{ij}\dfrac{\mathrm{d}q^{i}}{\mathrm{d}s}\dfrac{\mathrm{d}q^{j}}{\mathrm{d}s}}\mathrm{d}s \qquad \tilde{\gamma}(s_0)=a \quad \tilde{\gamma}(s_1)=b\,\,.
\end{equation}
One possible way to provide an identification of natural motion with geodesics is provided by the introduction of the Jacobi metric on a subspace of configuration space.
Let us consider the total energy function
\begin{equation}
H(\dot{q},q)=\left(\dot{q}^i\dfrac{\partial L}{\partial \dot{q}^i}-L\right)=\dfrac{1}{2} \overline{g}_{ij}(q) \dot{q}^{i} \dot{q}^j+V(q)
\end{equation}
which is obviously a conserved quantity along the natural motions, i.e. $d{H}(\dot{q}(t),q(t))/dt=0$. Since the Lagrangian is a homogeneous function in $\dot{q}^i$ it follows that
%\begin{equation}
$H(\dot{q},q)=2W-L$
%\end{equation}
where  
\begin{equation}
W=\dfrac{1}{2}g_{ij}\dot{q}^i\dot{q}^{j}=H(\dot{q},q)-V(q)
\end{equation}
is the kinetic energy.
If we consider a class of \textit{isoenergetic} trajectories $q(t;E)$ in configuration space,
that is having the same total energy value $H(\dot{q}(t;E),q(t;E))=E$, as the kinetic energy is non negative, the trajectories of the system in configuration space are confined in the region $\mathcal{M}_{V < E}=\left\{q\in M|V(q) < E\right\}$.
Moreover, for isoenergetic trajectories the kinetic energy can be expressed as a function of the coordinates, i.e.
\begin{equation}
W(q(t;E))=E-V(q)
\end{equation}
thus the action functional of Eq. \eqref{eq:action_functional} can be rewritten in the form
\begin{equation}
\begin{split}
\label{eq:reduced_action}
\mathit{S}_{E}[q(t;E)]=\int_{t_0}^{t_1} L(\dot{q}(t;E),q(t;E))\,\,\, \mathrm{d}t&=\int_{t_0}^{t_1}\left[E+2W(q(t;E))\right]\,\,\,\mathrm{d}t\\
&=(t_1-t_0)E+\int_{t_0}^{t_1}2W(q(t;E))\,\,\mathrm{d}t\,\,\,.
\end{split}
\end{equation}
as we are interested in the variational principle and $t_0,t_1,E$ are fixed quantities, the first term
in the last equality can be neglected.
The integral in Eq.\eqref{eq:reduced_action} can be interpreted as a length integral in configuration space, in fact
\begin{equation}
\label{eq:reduced_action_ds}
\begin{split}
\mathit{S}_{E}[q(t;E)]&=\int_{t_0}^{t_1}2W(q(t;E))\,\,\mathrm{d}t\\
&=\int_{t_0}^{t_1}\sqrt{2W(q(t;E))}\sqrt{2W(q(t;E))}\,\,\mathrm{d}t=\int_{t_0}^{t_1}\sqrt{2\left[E-V(q)\right]}\sqrt{\bar{g}_{ij}(q)\dot{q}^i\dot{q}^j}\,\,\mathrm{d}t=\\
&=\int_{t_0}^{t_1}\sqrt{2\left[E-V(q)\right]\bar{g}_{ij}\dot{q}^i\dot{q}^j}\,\,\mathrm{d}t=\int_{t_0}^{t_1}\sqrt{g_{ij}\dot{q}^i\dot{q}^j}\,\,\mathrm{d}t=\int_{t_0}^{t_1}\sqrt{\left(\dfrac{\mathrm{d}s}{\mathrm{d}t}\right)^2}\,\,\mathrm{d}t\\
&=\int_{s_0}^{s_1}\mathrm{d}s
\end{split}
\end{equation}
where the new metric
\begin{equation}
\label{eq:Jacobi_scaling_ch2}
g_{ij}:=2W(q) \overline{g}_{ij},\qquad W(q):=E-V(q)
\end{equation}
called also \textit{Jacobi metric}, has been introduced with the associated the arc-length element 
\begin{equation}
\mathrm{d}s^2=g_{ij}\mathrm{d}q^{i}\mathrm{d}q^j=2\left[E-V(q)\right]\overline{g}_{ij}\mathrm{d}q^{i}\mathrm{d}q^j=4[W(q)]^2\mathrm{d}t^2\,\,.
\end{equation}
A central point of the following discussion is related to the \textit{non-affine parametrization} of the arc-length  $s$ with respect to the physical time $t$, which  is clearly a necessary consequence of the construction of Jacobi metric.
Moreover, we observe that the Jacobi metric is related to kinetic energy metric $\overline{g}_{ij}$ through a \textit{conformal rescaling} via a factor proportional to the kinetic energy  $[E-V(q)]$ preserving the signature of the metric only in the interior of $\mathcal{M}_{V\leq E}$, i.e. $\mathcal{M}_{V < E}=\mathcal{M}_{E}=\left\{q\in M|V(q)<E\right\}$, the so called \textit{Hill's region}.
By endowing the region $\mathcal{M}_{E}$ with the Jacobi metric $g$ \textit{the natural motions with fixed energy $E$ are the same as 
geodesics $\gamma(s)$ of the manifold $(M_E,g)$}.
%As the factor $W(q)$ depends on the potential energy, the geometry (connection and curvature) of Hill's regions turns out non-trivial, especially for non-linear dynamical systems (since the second derivatives of the potential energy are not constant). 

This approach has remarkable consequences, as it is discussed in the following.
In fact, the geometric description of Newtonian dynamics, identifying the solutions of Newton equations with the geodesics of suitable Riemannian manifolds, 
provides a powerful conceptual and mathematical framework to study the stability/instability of dynamics in terms of the stability properties 
of a geodesic flow,  described by the geodesic spread equation relating stability/instability with geometry. 

The standard observable to define the presence of dynamical chaos and to measure its strength is the \emph{largest Lyapunov exponent}, and, as we will see in the following, in the geometrical framework a Geometrical Lyapunov exponent can be defined.
%Before to define it, we spent few words about the previous derivation and about what means to geometrize an Hamiltonian dynamical systems.

Now, a key point in derivation of Jacobi metric from Maupertuis' principle is to set the relation between the physical time $t$ and the arc-length $s$ as 
\begin{equation}
\label{eq:s_to_t_jacobi_ch2}
ds^{2}=4W^2(q) dt^{2}\,.
\end{equation}
It follows that a generic geodesic parametrized by the arc-length $s$, i.e. $\dot{\pieno{q}}(s)=\{\dot{q}^{i}(s)\}_{i\in[1,n]}$ has a unit velocity 
\begin{equation*}
g_{J}(\dot{q}(s),\dot{q}(s))=2W\frac{dq^{i}}{ds}\delta_{ij}\frac{dq^{j}}{ds}=1
\end{equation*}
while if parametrized with respect to the physical time
\begin{equation}
\label{eq:2W_of_dqdt}
\dfrac{dq^{i}}{dt}g_{ij}\frac{dq^{j}}{dt}=2W\,.
\end{equation}
So, in general, the physical time is a \textit{non-affine parametrization} of the geodesics of Jacobi metric.
We derive in what follows the effect of the reparametrization \eqref{eq:s_to_t_jacobi_ch2} on the geodesic equation. Let us introduce the vector fields 
\begin{equation}
Y^{i}:=\dfrac{\mathrm{d}q^i}{\mathrm{d}s}
\end{equation}
and
\begin{equation}
X^{i}:=\dfrac{\mathrm{d}\tilde{q}^i}{\mathrm{d} t}=\dfrac{\mathrm{d}s}{\mathrm{d}t}\dfrac{\mathrm{d}q^i}{\mathrm{d}s}=2W Y^{i}
\end{equation}
defined along the geodesic $q^{i}(s)=q^{i}(s(t))=\tilde{q}^{i}(t)$.
Using the definition of geodesic for the vector field $Y$
\begin{equation}
\nabla_{Y}Y=0
\end{equation}
it is possible to derive the equations for the vector field $X=\alpha Y$ (with $\alpha=(2W)^{-1}$)
\begin{equation}
\nabla_{\alpha X}\left(\alpha X\right)=\alpha \left(\nabla_{X}\alpha\right)+\alpha^2 X=0
\end{equation}
that implies
\begin{equation}
\label{eq:geod_dt_coorfree}
\nabla_{X}X=-\nabla_{X}(\log\alpha) X=X \nabla_{X}\left(\log 2W\right)\,.
\end{equation}
In a natural coordinate system $\{q^{i}\}_{i=1,...,N}$ the equation \eqref{eq:geod_dt_coorfree} reads
\begin{equation}
\label{eq:geod_dt_natcoor}
\frac{\mathrm{d}^{2}q^{i}}{\mathrm{d}t^{2}}+\Gamma^{i}_{jk}\frac{\mathrm{d}q^{j}}{dt}\frac{\mathrm{d}q^{k}}{dt}=\delta^{i}_{k}\dfrac{\mathrm{d}q^{k}}{\mathrm{d}t}\dfrac{\mathrm{d}q^{j}}{\mathrm{d}t}\partial_{j}\log(2W)\,.
\end{equation}
The Christoffel symbols in Jacobi metric take the form
\begin{equation}
\label{eq:ChristoffelJacobi}
\begin{split}
\Gamma^{i}_{jk}=&\overline{\Gamma}^{i}_{jk}+\dfrac{1}{2}\left[\delta^{i}_{j}\partial_{k}\log(2W)+\delta^{i}_{k}\partial_{j}\log(2W)-g_{jk}\overline{\mathrm{grad}}^{i}\log(2W)\right]
\end{split}
\end{equation}
where $\overline{\Gamma}^{i}_{jk}$ are Christoffel symbols of the kinetic energy metric $g_{ij}$ and $\overline{\mathrm{grad}}^{i}f:=g^{ik}\partial_{i}f$ is the gradient with respect to kinetic energy
metric. Substituting \eqref{eq:ChristoffelJacobi} in \eqref{eq:geod_dt_natcoor} and using 
\eqref{eq:2W_of_dqdt} we obtain 
\begin{equation}
\overline{\nabla}_{X}X=\frac{\mathrm{d}^{2}q^{i}}{\mathrm{d}t^{2}}+\overline{\Gamma}^{i}_{jk}\frac{\mathrm{d}q^{j}}{dt}\frac{\mathrm{d}q^{k}}{dt}=\dfrac{1}{2}\overline{\mathrm{grad}}^{i} (2 W)=-\overline{\mathrm{grad}}^{i} V
\end{equation}
i.e. the Newton's equations of the dynamical system.
As already mentioned above, dynamical chaos can now be investigated by means of the equation for the geodesic spread describing the stability of a geodesic flow of a Riemannian manifold. The geodesic spread is measured by a vector field $J$ which locally gives the distance between nearby geodesics. This vector field evolves along a reference geodesic according to the 
Jacobi-Levi Civita equation which, in components, reads
\begin{equation}
\label{eq:JLC_general_ch2}
\dfrac{\nabla^{2}J^{i}}{ds^{2}}+R^{i}_{jkl}\dfrac{dq^{j}}{ds}J^{k}\frac{dq^{l}}{ds}=0
\end{equation}
where 
\begin{equation}
R^{i}_{jkl}=\partial_{k}\Gamma^{i}_{jl}-\partial_{l}\Gamma^{i}_{jk}+\Gamma^{m}_{lj}\Gamma^{i}_{km}-\Gamma^{m}_{kj}\Gamma^{i}_{ml}
\end{equation}
is the Riemann curvature tensor.
In Jacobi metric we have
\begin{equation}
\begin{split}
\dfrac{\nabla^{2} X^{i}}{ds^{2}}&=\dfrac{1}{2W}\dfrac{\nabla}{dt}\left(\frac{1}{2W}\dfrac{\nabla X^{i}}{dt}\right)\\
&=\dfrac{1}{4W^{2}}\dfrac{\nabla^{2}X^{i}}{dt^{2}}+\dfrac{1}{4W}\dfrac{\nabla X^{i}}{dt}\dfrac{d}{dt}\left(\dfrac{1}{W}\right)
\end{split}
\end{equation}
that substituted into \eqref{eq:JLC_general_ch2} yields to
\begin{equation}
\label{eq:JLC_jacobi_repar}
\begin{split}
\frac{\nabla^{2}J^{i}}{ds^{2}}+R^{i}_{jkl}\frac{dq^{j}}{ds}J^{k}\frac{dq^{l}}{ds}=0\mapsto\frac{\nabla^{2}J^{i}}{dt^{2}}+R^{i}_{jkl}\frac{dq^{j}}{dt}J^{k}\frac{dq^{l}}{dt}=\frac{\nabla J^{i}}{dt}\frac{d}{dt}\ln W\,.
\end{split}
\end{equation}
Every non-affine parametrization in Jacobi metric can be reduced to a relation between the physical time and the arc-length parameter of the form $t\mapsto ds=f(q)dt$, where $f(q)$ is a function of the point and  generates a term proportional to the derivative of $f(q)$ into each equations.
Thus, the JLC equation written for the Jacobi metric is not invariant for time reparametrization. We can rephrase the main point raised by the work in Ref.\cite{cuervo2015non} as attributing to the right hand-side of Eq.\eqref{eq:JLC_jacobi_repar} the origin of  chaotic-like instabilities even for  integrable systems, that is, the origin of 
non-physical artifacts. In Ref. \cite{cuervo2015non} it has been surmised that the larger the fluctuations of kinetic energy $W$ the more dramatic  the occurrence of non-physical instabilities stemming from Eq.\eqref{eq:JLC_jacobi_repar}. 
Remarkably, the geometrization of Newtonian dynamics in an enlarged configuration space-time equipped with the Eisenhart metric tensor  \cite{marco} yields the JLC equation in the form of the Tangent Dynamics Equation which is commonly used to compute the Largest Lyapunov Exponent (LLE) \cite{marco,book}, therefore the authors of Ref. \cite{cuervo2015non} claim that this is the only consistent geometrization of Newtonian dynamics to investigate chaos, while the Jacobi metric would be unsuitable for the same task. 
%%%%%%%%%%%%%%%%%%%%%%%%%%%%%%%%%%%%%%%%%%%%%%%%%%%%%%%%%%%%%%%%%%%
\subsection{Geometrical Lyapunov exponent}
We conclude the present Section by giving a definition of a Geometrical Lyapunov exponent. Of course, the starting point is  the JLC which, in intrinsic notation, is
\begin{equation}
\begin{split}
\nabla_{\pieno{\xi}}^{2}J+R(J,\pieno{\xi})\pieno{\xi}=0 \ .
\end{split}
\end{equation}
By defining $Y=\nabla_{\nero{\xi}}J$ and $R_{\pieno{\xi}}J:=R(\pieno{\xi},J)\pieno{\xi}$  the above equation becomes
\begin{equation}
\begin{split}
\nabla_{\nero{\xi}}J&=Y\\
\nabla_{\pieno{\xi}}Y&=R_{\pieno{\xi}}J \ .
\end{split}
\end{equation}
Then, by putting 
\[
\mathcal{A}:=
\left(
\begin{matrix}
0 && 1\!\!1 \\
R_{\pieno{\xi}} && 0
\end{matrix}
\right),\qquad\qquad \mathcal{J}:=
\left(
\begin{matrix}
J  \\
Y
\end{matrix}
\right)
\]
the JLC equation reads
\begin{equation}
\label{compact_JLC}
    \begin{split}
        \nabla_{\nero{\xi}}\mathcal{J}=\mathcal{A}\mathcal{J} \ .
    \end{split}
\end{equation}
A Geometrical Lyapunov exponent, in analogy with the definition of the standard Lyapunov exponent, can be defined after having expressed as a function of physical time the solution of Eq. \eqref{compact_JLC} as
%\begin{eqnarray}
%\label{geometrical_lyapunov_exponent}
%\lambda_{G}:=\lim_{t\longrightarrow\infty}\frac{1}{t} \frac{\|\mathcal{J}(t)\|_{\pieno{g}^{S}}}{\|\mathcal{J}(0)\|_{\pieno{g}^{S}}} \ .
%\end{eqnarray}
%Explicitly, the norm of $\mathcal{J}$ is given by
%\[
%\|\mathcal{J}(t)\|^{2}_{\pieno{g}^{S}}=g^{S}_{ij}\mathcal{J}^{i}\mathcal{J}^{j}
%\]
%where $\pieno{g}^{S}$  is the metric tensor obtained from the Sasaki lift of the Jacobi metric (see \cite{book}).
\begin{eqnarray}
\label{geometrical_lyapunov_exponent}
\lambda_{G}:=\lim_{t\longrightarrow\infty}\frac{1}{t} \frac{\|\mathcal{J}(t)\|_{\pieno{g}}}{\|\mathcal{J}(0)\|_{\pieno{g}}} \ .
\end{eqnarray}
where the norm of $\mathcal{J}$ is
\[
\|\mathcal{J}(t)\|^{2}_{\pieno{g}}=2\,W(\pieno{q})\delta_{ij}\{J^{i}(t)J^{j}(t)+(\nabla_{\dot{\pieno{q}}}J)^{i}(t)(\nabla_{\dot{\pieno{q}}}J)^{j}(t)\} \ .
\]
Let us note that throughout the literature, \cite{marco}, \cite{book}, \cite{cerruti1996geometric} and \cite{cerruti1997lyapunov}, the geometrical Lyapunov exponent has been expressed as a function of physical time $t$, whereas in \cite{cuervo2015non}  the Geometrical Lyapunov exponent is defined as a function of the arc-length.
As a final comment, let us remark that the existence of many different frameworks to rephrase Newtonian dynamics in geometric terms \cite{book}
can lead, a-priori, to different \textit{quantitative} evaluations of the strength of chaos, to the contrary, the use of different geometric frameworks must lead to the same \textit{qualitative} description of the stability/instability properties of the dynamics. For example, the transition from weak to strong chaos must be at least qualitatively and coherently reproduced in any geometric framework, as is for example actually shown for high dimensional Hamiltonian flows in Ref. \cite{cerruti1997lyapunov}. 

%%%%%%%%%%%%%%%%%%%%%%%%%%%%%%%%%%%%%%%%%%%%%%%%%%%%%%%%%%%%%%%%%%%
\section{Parallel Transported Frame for a system of $N$ particles in Jacobi Manifold}
\label{sec_parallel_transported_frame}
Let us now work out the JLC equation for a parallel transported orthonormal frame along a reference geodesic. The advantage of this representation with respect to the use of standard local coordinates is that by making  parallel transported frames to "incorporate" the geodesics reflection, when they approach the Hill's boundaries in configuration space, eliminates a source of artefacts in the numerical solution of the JLC equation. In fact, the sharp reflection of geodesics close to the Hill's boundary of a mechanical manifold would require a prohibitively high numerical precision to avoid the introduction of an error amplification mimicking chaos even for integrable systems. To the contrary, with respect to a parallel transported frame - "incorporating" the geodesic reflection - nearby geodesics are no longer affected by the fake error amplification due to the reflection.
As a matter of fact, we will show that  the solutions of the JLC equation - written for the Jacobi metric - have the correct physical meaning also in the "pathological" cases where unphysical instabilities were found in Ref.\cite{cuervo2015non}. 

The \emph{parallel transported frame} is built by requiring that the covariant derivative $\nabla_{\nero{\xi}}X$ of all vectors $X$ with respect to the geodesic flow  $\nero{\xi}$ is orthogonal to $\nero{\xi}$. 
Let us introduce  a reference frame $e_{i}:=\partial/\partial q^{i}$ on the tangent bundle $TM$ and the corresponding \emph{dual frame} $\theta^{i}:=dq^{i}$ such that $\theta^{i}(e_{j})=\delta^{i}_{j}$.
The Jacobi metric tensor is the multi-linear map $\pieno{g}_{J}:TM\times TM\longrightarrow{\Bbb R}$ such that, if  written with respect to the \emph{natural basis}, is given by
\begin{equation}
\pieno{g}_{J}:=g_{ij}\theta^{i}\otimes \theta^{j}
\end{equation}
the components of which are 
\begin{equation*}
g_{ij}:=\pieno{g}(e_{i},e_{j})=2(E-V(q))\delta_{ij} 
\end{equation*}
with the differential arc-length
\begin{equation}
ds^{2}=g_{ij}\theta^{i}\theta^{j} \ .
\end{equation}
To build a \emph{parallel transported frame} we consider a geodesic $\gamma:I\subset{\Bbb R}\longrightarrow M$, 
with tangent vector field $\nero{\xi}:I\longrightarrow TM$ and the Jacobi vector field $J:I\longrightarrow TM$
such that respect to the natural basis  $\{e_{i}\}_{i\in[1,n]}$ they are written as
\begin{equation*}
\begin{split}
J&=J^{i}e_{i}\\
\nero{\xi}&=\xi^{i}e_{i}
\end{split}
\end{equation*}
where the Jacobi field, by definition, verifies
\begin{equation}
\nabla_{\nero{\xi}}^{2}J+R(J,\nero{\xi})\nero{\xi}=0
\label{JLC_chapter2}
\end{equation}
We shall show that there exist  reference frames parallel transported along any geodesic.
Such systems exist if an orthogonal tensor field $\nero{\Omega}:I\longrightarrow SO_{N}(\gamma(I))$  is defined at each point along the geodesic flow. This tensor field allows to define the required frame where the basis on $TM$ is given by $\{E_{i}\}_{i=i,\cdots,N}$ and dual frame on $T^{*}M$ $\{\Theta^{i}\}_{i=i,\cdots,N}$ and, by definition of dual frame we have $\Theta^{i}(E_{j})=\delta^{i}_{j}$.
Therefore, the \emph{parallel transported frame} is defined by
\begin{equation}\label{def_Omega}
\begin{split}
E_{i}=\nero{\Omega}\cdot e_{i},\quad \nabla_{\nero{\xi}}E_{i}&=0,\quad \pieno{g}(E_{i},E_{j})=\delta_{ij}\\
\Theta^{i}=\theta^{i}\cdot\nero{\Omega}^{-1},\quad \nabla_{\nero{\xi}}\Theta^{i}&=0,\quad \pieno{g}^{*}(\Theta^{i},\Theta^{j})=\delta^{ij}
\end{split}
\end{equation} 
In the Jacobi-Levi Civita equation, written in \eqref{JLC_chapter2}, we can define the Ricci tensor along the geodesic flow $Ric(\nero{\xi}):\gamma(I)\longrightarrow \mathcal{T}^{1}_{1}(\gamma(I))$  by
\begin{equation}
Ric(\nero{\xi}):=R(\cdot,\nero{\xi})\nero{\xi} \ .
\end{equation}
A canonical isomorphism exists between the rank-two tensor $Ric(\nero{\xi})$ and a  symmetric matrix $N\times N$, that we shall denote with $\nero{R}$. For every $\omega\in T^{*}M$ and $X\in TM$, such a matrix is given by
\begin{equation}
\begin{split}
Ric(\nero{\xi})(X,\omega)&:=Ric(\nero{\xi})^{i}\,_{j}X^{j}\omega_{i}=\pieno{g}(Y_{\omega},\nero{R}(X))
\end{split}
\end{equation}
where $Y_{\omega}:=\pieno{g}^{-1}\omega$.
We note the following symmetry properties of $Ric(\nero{\xi})$ 
\begin{equation}
\begin{split}
Ric(\nero{\xi})&=R_{jkl}\,^{i}\xi^{k}\xi^{l} e_{i}\otimes\theta^{j}=R_{jkli}\xi^{k}\xi^{l} \theta^{i}\otimes\theta^{j}\\
&=R_{ilkj}\xi^{k}\xi^{l} \theta^{i}\otimes\theta^{j}=R_{ilk}\,^{j}\xi^{k}\xi^{j} e_{j}\otimes\theta^{i}
\end{split}
\end{equation}
and by redefining the indices $j\rightleftharpoons i$ and $l\rightleftharpoons k$ the symmetry of  $Ric(\nero{\xi})$  is evident. 
Hence the associated matrix can be diagonalized at each point along the flow.
Let note that the symmetry of Riemann tensor entails
\begin{equation}
Ric(\nero{\xi})\nero{\xi}=R(\nero{\xi},\nero{\xi})\nero{\xi}=0 \ .
\end{equation}
Then, the tangent vector to the geodesic $\nero{\xi}$ is an eigenvector of the matrix $\nero{R}$ associated to $Ric(\nero{\xi})$ with vanishing eigenvalue.
A posteriori, one observes that for harmonic oscillators - geometrized through the Jacobi metric - the eigenbasis of the matrix $\nero{R}$ coincides with the parallel transported frame. In general, this is not true and we have two different orthogonal matrices, that one which diagonalises $\nero{R}$ and another one which transforms the natural basis into the parallel transported frame.
Fortunately, in the present case, given the natural frame $\{e_{i}\}_{i\in[1,N]}$ on the tangent bundle $TM$, there exists an orthogonal transformation $\nero{\Omega}:I\longrightarrow SO_{N}(\gamma(I))$ such that \begin{equation}
E_{i}=\nero{\Omega}\cdot e_{i}
\end{equation}
namely, such that it transforms the natural basis into the parallel transported frame $\{E_{i}\}_{i\in[1,N]}$ and, moreover, such that it diagonalises the matrix $\nero{R}$:
\begin{equation}
\nero{R}_{d}=\nero{\Omega}^{-1}\circ \nero{R}\circ\nero{\Omega}
\end{equation}
This allows to write the Jacobi field with respect to such a basis and thus
\begin{equation}
\begin{split}
J=J^{i}e_{i}&=J^{i}(\nero{\Omega}^{-1})^{k}_{i}\nero{\Omega}^{l}_{k}e_{l}\\
&=[(\nero{\Omega}^{-1})^{k}_{i}J^{i}][\nero{\Omega}^{l}_{k}e_{l}]=\tilde{J}^{k}E_{k} \ .
\end{split}
\end{equation}
Let us consider the JLC equation \eqref{JLC_chapter2} and proceed to substitute $J\mapsto\tilde{J}$ then
\begin{equation}
\begin{split}
&\nabla_{\nero{\xi}}^{2}\tilde{J}+Ric(\nero{\xi})\tilde{J}=0\\
&\frac{d^{2}\tilde{J}^{i}}{ds^{2}}E_{i}+\tilde{J}^{k}Ric(\nero{\xi})^{l}\,_{k}E_{l}=0\\
&\left(\frac{d^{2}\tilde{J}^{k}}{ds^{2}}+\tilde{J}^{k}\Lambda_{k}\right)E_{k}=0\\
\end{split}
\end{equation}
where now the repeated indices do not stand for summation. 
In this way, we obtain $N -1$ second order differential equations in the unknown functions $\{\tilde{J}^{k}\}_{i\in[1,N-1]}$, and the $N$-th equation is that for the vector tangent to the reference geodesic, equation which corresponds to the geodesic equation thus, by denoting with $\tilde{J}^{1}$ the function for this equation, we have
\begin{equation}
\begin{split}
&\frac{d^{2}\tilde{J}^{1}}{ds^{2}}=0\\
&\frac{d^{2}\tilde{J}^{k}}{ds^{2}}+\tilde{J}^{k}\Lambda_{k}=0
\label{JLC trasportata par. ogni metrica}
\end{split}
\end{equation}
It is interesting to remark that $\{\Lambda_{k}\}$ are just \emph{sectional curvatures}, i.e the principal directions of curvature identified by the vectors $\{E_{k}\}_{k\in[1,N]}$, i.e
\begin{equation*}
\Lambda_{k}:=\langle E_{k},R(E_{k},E_{1})E_{1}\rangle
\end{equation*}
Now, passing from the arc-length parameter $s$ to the physical time $t$  the equations \eqref{JLC trasportata par. ogni metrica} become
\begin{equation}
\frac{d^{2}\tilde{J}^{k}}{dt^{2}}(t)-\frac{d}{dt}\log(W(q(t)))\frac{d\tilde{J}^{k}}{dt}(t)+4W(q(t))^{2}\tilde{J}^{k}(t)\Lambda_{k}(t)=0
\end{equation}
whence
\begin{equation}
\begin{split}
\label{before_canonical_way}
&\ddot{J}^{k}(t)-a(t)\dot{J}^{k}(t)+b(t)J^{k}(t)=0 \qquad k\in[2,N]\\
&\ddot{J}^{1}(t)-a(t)\dot{J}^{1}(t)=0\\
a(t)&:=\frac{d}{dt}\log(W(q(t)))\\
b_{k}(t)&:=4W(q(t))^{2}\Lambda_{k} \ .
\end{split}
\end{equation}
The second equation in \ref{before_canonical_way} can be immediately solved by setting $h(t):=\dot{J}^{1}(t)$ and then tackling the differential equation
\begin{equation*}
\dot{h}(t)+a(t)h(t)=0\implies h(t)=W(q(t))
\end{equation*}
giving
\begin{equation}
J^{1}(t)=\int^{t}_{t_{0}} W(q(\eta))\,d\eta \ .
\end{equation}
The other equations can be written in \emph{canonical form}, namely, by redefining the function $\tilde{J}$ as follows
\begin{equation}
\tilde{J}=A(t) f(t)
\end{equation}
and substituting it into \eqref{before_canonical_way} we obtain 
\begin{equation}
\label{Eqeffe}
\ddot{f}(t)+\dot{f}(t)\left(-a(t)+2\frac{\dot{A}(t)}{A(t)}\right)+f(t)\left(-a(t)\frac{\dot{A}(t)}{A(t)}+\frac{\ddot{A}(t)}{A(t)}+b(t)\right)=0 \ .
\end{equation}
By choosing $A(t)$ such that the coefficient of $\dot f(t)$ vanishes, we get the following conditions for $A(t)$
\begin{equation}
\begin{split}
\frac{\dot{A}}{A}&=\frac{a(t)}{2}\\
\frac{\ddot{A}}{A}&=\frac{a(t)^{2}}{4}+\frac{\dot{a}(t)}{2}
\end{split}
\end{equation}
thus giving 
\begin{equation}
\begin{split}
A(t)&=\exp\left(\frac{1}{2}\int a(t) \;dt\right)=\exp\left(\frac{1}{2}\int\frac{d}{dt}\log(W(q(t))) \;dt\right)\\
&=\exp\left(\frac{1}{2}\log(W(q(t)))\right)=\sqrt{W(q(t))} \ .
\end{split}
\end{equation}
The equation \eqref{Eqeffe} becomes
\begin{equation}
\begin{split}
\ddot{f}(t)+f(t)\left(-\frac{a(t)^{2}}{4}+\frac{\ddot{a}(t)}{2}+b(t)\right)=0\\
\ddot{f}(t)+f(t)\left(-\frac{3}{4}\left(\frac{\dot{W}}{W}\right)^{2}+\frac{1}{2}\frac{\ddot{W}}{W}+b(t)\right)=0
\end{split}
\end{equation}
To apply this procedure to the equations \eqref{JLC trasportata par. ogni metrica}, we set 
\begin{equation}\label{forma canonica}
\tilde{J}^{k}:=A(t)f^{k}(t)
\end{equation}
and for every $k\in[2,N]$ we obtain the final form for the components of the Jacobi-Levi Civita equation
\begin{equation}
\ddot{f}^{k}+\omega_{(k)}(t)f^{k}(t)=0
\end{equation}
where the functions in the second term of the l.h.s. can be interpreted as time dependent frequencies 
\begin{equation}
\omega_{(k)}(t) =4W(q(t))^{2}\Lambda_{k}(t)-\frac{3}{4}\left(\frac{\dot{W}}{W}\right)^{2}+\frac{1}{2}\frac{\ddot{W}}{W}
\end{equation}
where
\begin{equation}
\begin{split}
\dot{W} = -\dot{V}&= - \langle\nero{\xi},\nabla^{{\Bbb R}^{n}}V\rangle_{{\Bbb R}^{n}}\\
\ddot{W} = -\ddot{V}&= - HessV(\nero{\xi},\nero{\xi}) + \|\nabla^{{\Bbb R}^{n}} V\|_{{\Bbb R}^{n}}^{2}
\end{split}
\end{equation}
\subsection{Parallel Transported Frame for a system of $2$ harmonic oscillators}
 \label{subsec_simulation_parallel_transported_frame2} 
In Refs. \cite{cerruti1996geometric,rick} it has been shown that for the Hénon-Heiles model and for two coupled quartic oscillators, respectively, the geometrization through the Jacobi metric perfectly discriminates between ordered and chaotic motions by investigating the stability/instability of geodesics through the JLC equation expressed in a parallel transported frame. In this section, we are going to show that for two harmonic oscillators (of course an integrable system) the norm of the geodesic separation vector remains bounded, in spite of the fluctuations of kinetic energy which, according to the claim of Ref.\cite{cuervo2015non}, should have entailed apparent instability of the regular motions of this system.
The Hamiltonian of this system is
\begin{equation}
H(p_{1},p_{2},q_{1},q_{2}) =\frac{1}{2}(p_{1}^{2}+p_{2}^{2})+\frac{\kappa}{2}(q_{1}^{2}+q_{2}^{2})
\end{equation}
and the associated Jacobi metric, having set  $W(q):=E-V(q_{1},q_{2})$, is 
\begin{equation}
\pieno{g}_{J} =2W(q)(dq^{1}\otimes dq^{1}+dq^{2}\otimes dq^{2}) \ .
\end{equation}
The Ricci tensor along the geodesic flow is
\begin{equation}
[Ric(\nero{\xi})]^{i}\,_{j}=\frac{E\kappa}{W^{2}}\left(\begin{matrix}
\xi^{2}\xi^{2} && -\xi^{1}\xi^{2}\\
-\xi^{1}\xi^{2} && \xi^{1}\xi^{1} \ .
\end{matrix}\right)
\end{equation}
The eigenvectors of this matrix are
\begin{equation}
\begin{split}
E_{1}& =(\xi^{1},\xi^{2})^{t}\\ E_{2}&=(-\xi^{2},\xi^{1})^{t}
\end{split}
\end{equation}
with the corresponding eigenvalues:
\begin{equation}
\begin{split}
\Lambda_{1}&=0\\
\Lambda_{2}&=\frac{E \kappa}{2W^{3}} \ .
\end{split}
\end{equation}
With these eigenvectors the matrix for the basis transformation is simply obtained in the form
\begin{equation}
\nero{\Omega} =\left(\begin{matrix}
\xi^{1} && -\xi^{2}\\
\xi^{2} && \xi^{1}
\end{matrix}\right)
\end{equation}
so that the JLC equation for the components of the parallel transported Jacobi vector field $\tilde{J}$ is written as
\begin{equation}
\begin{split}
&\frac{d^{2}\tilde{J}^{1}}{ds^{2}}=0\\
&\frac{d^{2}\tilde{J}^{2}}{ds^{2}}+\frac{E\kappa}{2W^{3}}\tilde{J}^{2}=0
\end{split}
\end{equation}
which, written for the physical time $t$ and with the notations of the preceding Section, read 
\begin{equation}\label{JLC Jacobi 2 dimensioni nel tempo}
\begin{split}
&\frac{d^{2}J^{1}}{dt^{2}}-\frac{\dot{W}}{W}\frac{dJ^{1}}{dt}=0\\
&\frac{d^{2}f^{2}}{dt^{2}}+\tilde{\omega}_{2}f^{2}=0
\end{split}
\end{equation}
where
\begin{equation}
\begin{split}
\tilde{\omega}_{2}(t) =\frac{2E \kappa}{W}-\frac{3}{4}\left(\frac{\dot{W}}{W}\right)^{2}+\frac{1}{2}\frac{\ddot{W}}{W}\\
\label{frequency_jacobimetric_2dimension}
\end{split}
\end{equation}
\subsection{Numerical results}
\label{simulation2}
In Ref.\cite{cuervo2015non}, it has been claimed that there exists a set of initial conditions for which the JLC equations written for the Jacobi metric lead to unstable solutions even in the case of two harmonic oscillators because of the affine parametrization of the arc-length with time and the consequent fluctuations of the kinetic energy.
By starting from the general solutions for a system of two harmonic oscillators given by
\begin{equation}
\begin{split}
\label{q1q2}
q_{1}(t)&=A_{1}\cos(\omega t+\phi_{1})\\
q_{2}(t)&=A_{2}\cos(\omega t+\phi_{2})
\end{split}
\end{equation}
where $A_{1},A_{2}$ and $\phi_{1},\phi_{2}$ are determined by the initial conditions, the authors of Ref.\cite{cuervo2015non} used polar coordinates to rewrite the previous equations  in the compact form 
\begin{equation*}
\begin{split}
r(t)&=R^{2}+\Delta^{2}\cos(2\omega\,t)
\end{split}
\end{equation*}
where
\begin{equation}
R^{2}=\frac{A_{1}^{2}+A_{2}^{2}}{2},\quad \Delta^{2}=\frac{A_{1}^{2}-A_{2}^{2}}{2} \ ,
\end{equation}
then they have reported that every initial condition fulfilling  the condition
\begin{equation}\label{cuervo-reyes_condition}
I = \left(\frac{\Delta}{R}\right)^{4}> \frac{4}{7}
\end{equation}
yields unstable solutions.
We have adopted the same initial conditions for two harmonic oscillators and representing the solutions of the JLC equations with respect to a parallel transported frame, already used in \cite{cerruti1996geometric}, and we have found that the equations \eqref{JLC Jacobi 2 dimensioni nel tempo} with the frequencies \eqref{frequency_jacobimetric_2dimension} display stable solutions.
We have solved the equations by using two different conditions both fulfilling Eq. \eqref{cuervo-reyes_condition}. These are 
\begin{eqnarray}
I_{0}=\left(\frac{\Delta}{R}\right)^{4}&=&\frac{56}{81} \qquad \label{condition_0}\\
I_{1}=\left(\frac{\Delta}{R}\right)^{4}&=&\frac{21}{25} \qquad
\label{condition_1}
\end{eqnarray}

\begin{figure}[h!]
	\centering
	\includegraphics[height=8cm, width=9cm,keepaspectratio]{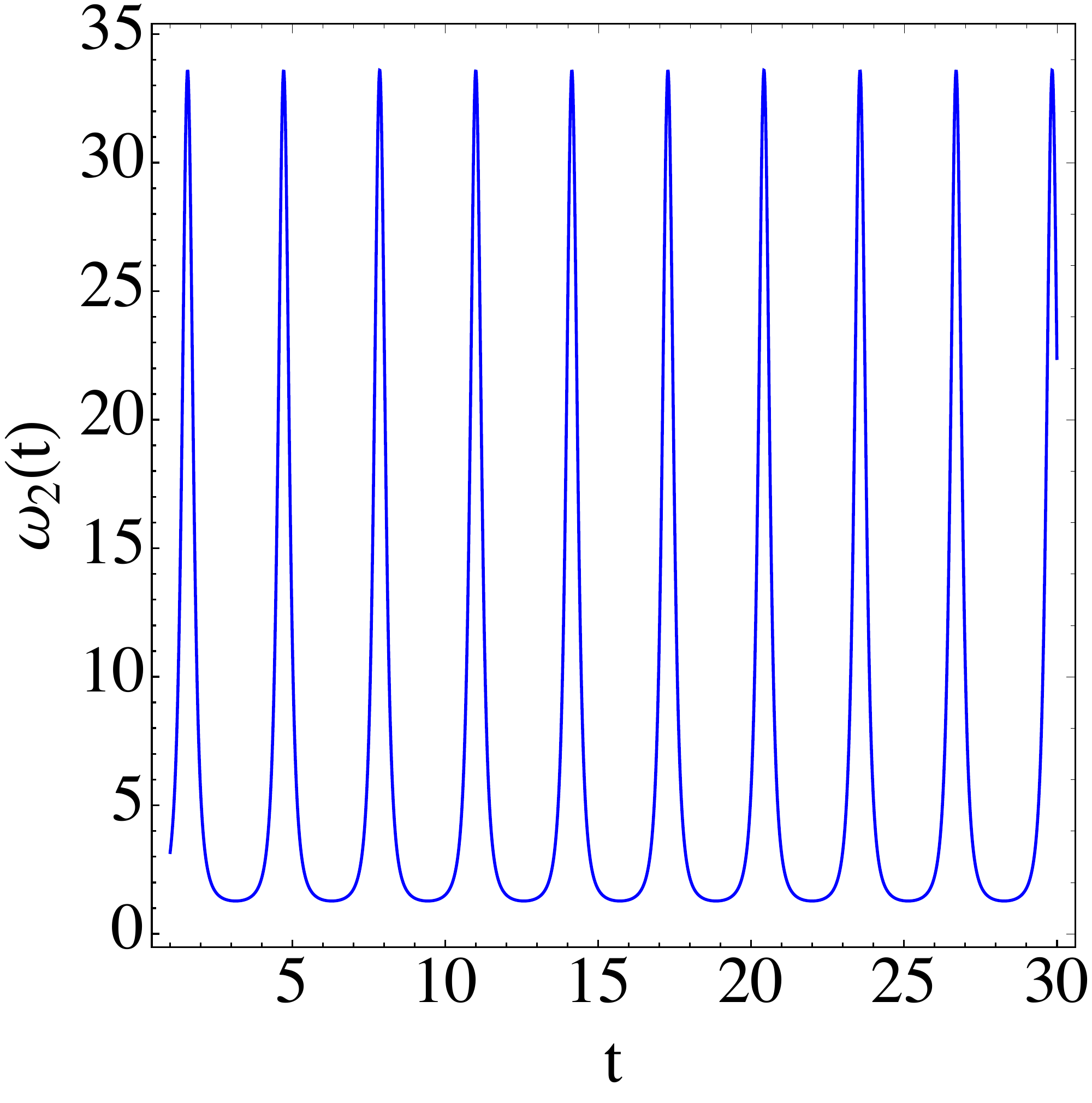}
	\caption{Time dependence of the function $\tilde{\omega}_{2}(t)$ defined in \protect\eqref{frequency_jacobimetric_2dimension}, for the condition \eqref{condition_0}.}
	\label{figura1}
	\end{figure}

\begin{figure}[h!]
	\centering
	\includegraphics[height=8cm, width=9cm,keepaspectratio]{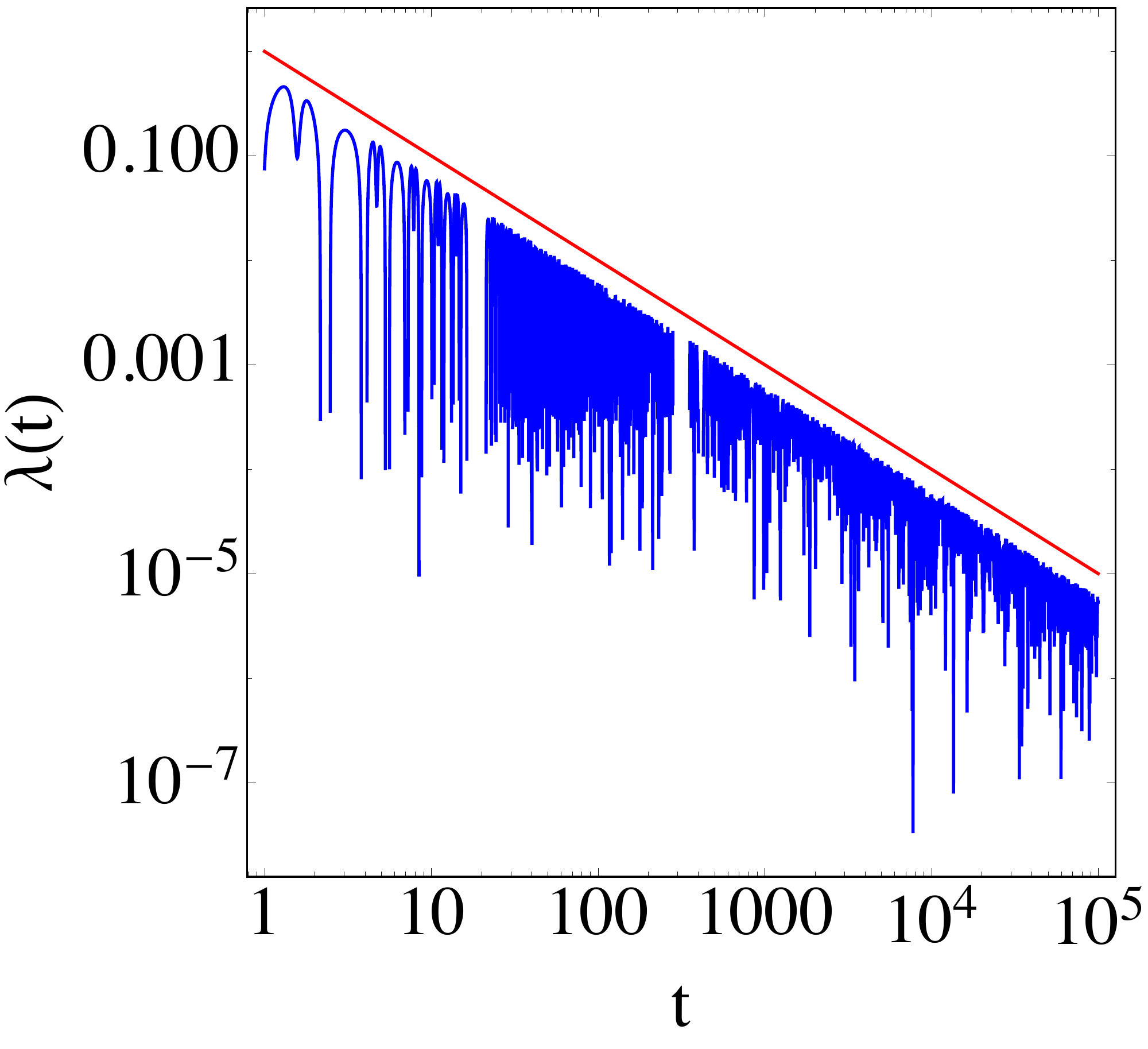}
	\caption{Time dependence of the Geometric Lyapunov Exponent $\lambda(t)$ associated to the $f^{2}(t)$ solution of Eqs.\eqref{JLC Jacobi 2 dimensioni nel tempo}, for the condition \eqref{condition_0}.
	Log-Log scale is used to evidence the $1/t$ decay represented by the red line.}
	\label{2oscillatori_jlc_jacobi_exp_lyap_geo_I1condizione}
\end{figure}

Let us consider the condition \eqref{condition_0}.
This is obtained by plugging through Eqs.\eqref{q1q2}  the initial condition 
\begin{equation}
A_{1}\to \frac{1}{2},\; A_{2}\to \frac{1}{10} \left(9+2 \sqrt{14}\right),\; \phi_{1}\to 0,\; \phi_{2}\to \frac{\pi}{2}
\end{equation}
into the frequency \eqref{frequency_jacobimetric_2dimension}. The time variation of $\tilde{\omega}_{2}(t)$ is reported in Figure \ref{figura1}. 
%and \ref{2oscillatori_jlc_jacobi_omega2_tilde_I1condizione}.
%\vskip 2truecm
%\begin{figure}[h!]	
%	\includegraphics[height=7cm, width=7cm,keepaspectratio]{omega2zoom_2dim_cond81.pdf}
%	\caption{Zoom of the frequency arised from \eqref{frequency_jacobimetric_2dimension}.}
%	\label{2oscillatori_jlc_jacobi_omega2_tilde_I1condizione}
%\end{figure}
Correspondingly, the time dependence of the Geometric Lyapunov Exponent, $\lambda(t)$, reported in Figure \ref{2oscillatori_jlc_jacobi_exp_lyap_geo_I1condizione}, clearly displays the typical time dependence found for regular trajectories, that is, $\lambda(t)\sim 1/t$.

Coming now to the second condition given in \eqref{condition_1}, again this is implemented by plugging through Eqs.\eqref{q1q2}  the initial condition 
\begin{equation}
A_{1}\to \frac{1}{2},\; A_{2}\to \frac{1}{4} \left(-5+\sqrt{21}\right),\; \phi_{1}\to 0,\; \phi_{2}\to \frac{\pi}{3}
\end{equation}
into the frequency \eqref{frequency_jacobimetric_2dimension}. 

\begin{figure}[h!]
	\centering
	\includegraphics[height=8cm, width=9cm,keepaspectratio]{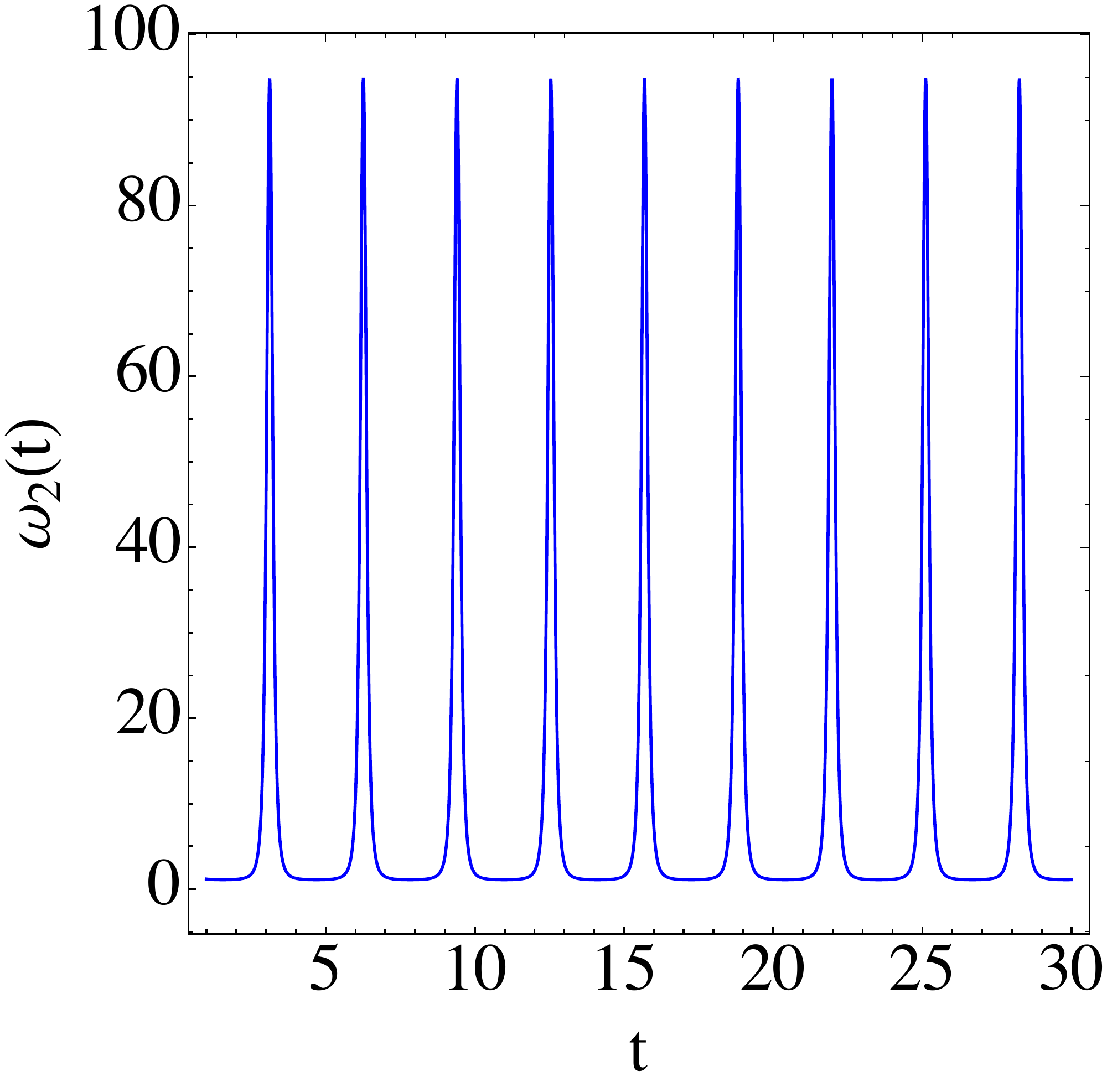}
	\caption{Time dependence of the function $\tilde{\omega}_{2}(t)$ defined in \protect\eqref{frequency_jacobimetric_2dimension}, for the condition \eqref{condition_1}.}
	\label{figura3}
%\centering
%	\includegraphics[height=7cm, width=7cm,keepaspectratio]{omega2zoom_2dim_cond83.pdf}
%	\caption{Zoom of the frequency which arises from \eqref{frequency_jacobimetric_2dimension} for the condition \eqref{condition_1}.}
%	\label{2oscillatori_jlc_jacobi_omega2_tilde_I3condizione}
\end{figure}

\begin{figure}[h!]
	\centering
	\includegraphics[height=8cm, width=9cm,keepaspectratio]{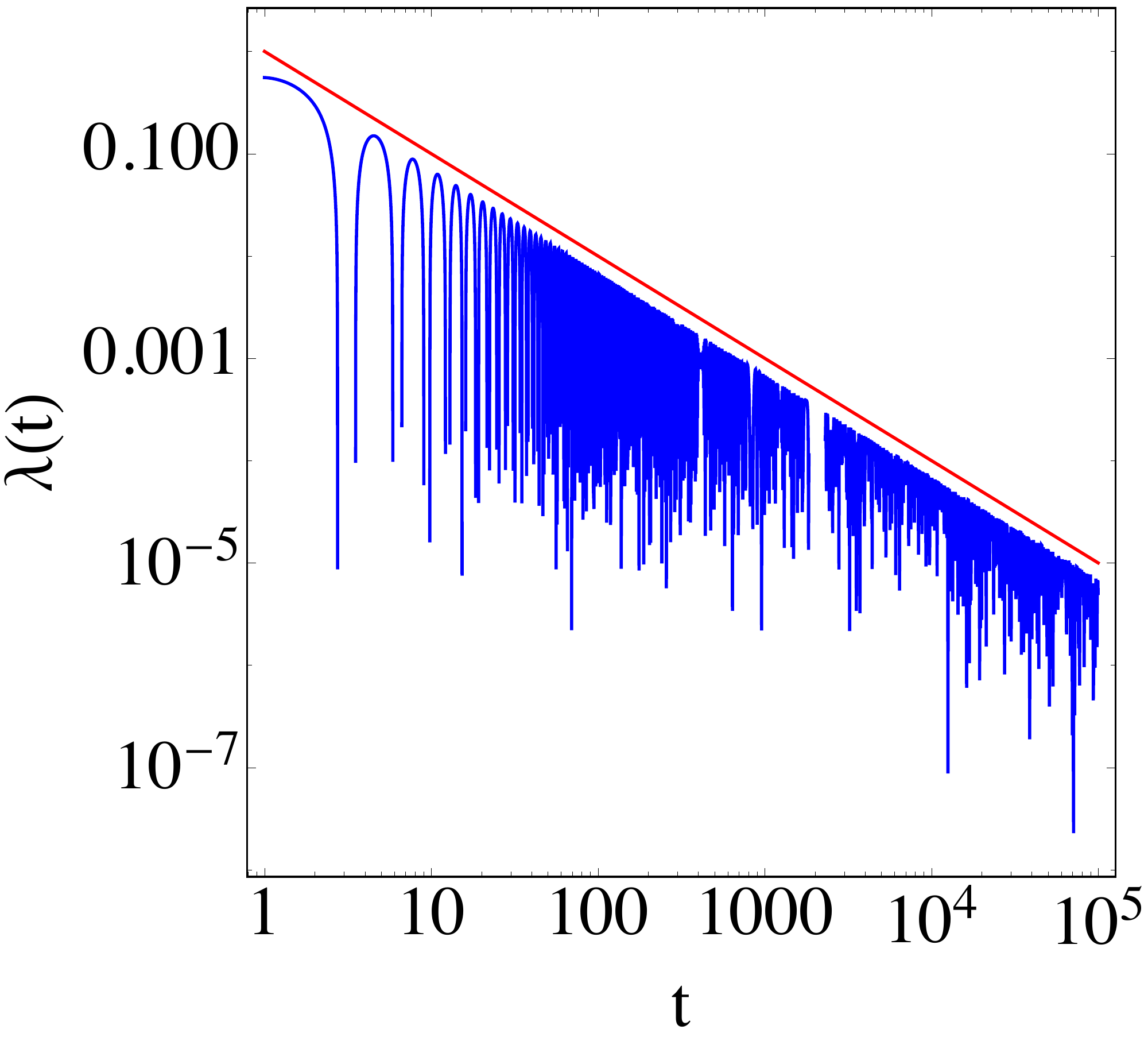}
	\caption{Time dependence of the Geometric Lyapunov Exponent $\lambda(t)$ associated to the $f^{2}(t)$ solution of Eqs.\eqref{JLC Jacobi 2 dimensioni nel tempo}, for the condition \eqref{condition_1}.
	Log-Log scale is used to evidence the $1/t$ decay represented by the red line.}
	 \label{2oscillatori_jlc_jacobi_omega2_tilde_I3condizione}
\end{figure}

The time variation of $\tilde{\omega}_{2}(t)$ is now reported in Figure \ref{figura3}. The corresponding time variation of the Geometric Lyapunov Exponent, $\lambda(t)$, is now reported in Figure \ref{2oscillatori_jlc_jacobi_omega2_tilde_I3condizione}. Again it is found that $\lambda(t)$ decays as $1/t$, as expected for regular motions.

%\pagebreak[3]
\clearpage

\subsection{Parallel Transported Frame for a system of $3$ harmonic oscillators}
 \label{subsec_simulation_parallel_transported_frame3}
A relevant step forward is now obtained by considering three degrees of freedom because now the geodesic separation vector has two nontrivial components (since the component parallel to the velocity vector does not accelerate). Therefore, we  consider $3$ harmonic oscillators described by the Hamiltonian 
\begin{equation}
H(p_{1},p_{2},p_{3},q_{1},q_{2},q_{3})=\frac{1}{2}(p_{1}^{2}+p_{2}^{2}+p_{3}^{2})+\frac{\kappa}{2}(q_{1}^{2}+q_{2}^{2}+q_{3}^{2})
\end{equation}
and the corresponding Jacobi metric, again with $W(q):=E-V(q_{1},q_{2},q_{3})$, is
\begin{equation}
\pieno{g}_{J}=2W(q)(dq^{1}\otimes dq^{1}+dq^{2}\otimes dq^{2}+dq^{3}\otimes dq^{3})
\end{equation}
The eigenvectors of the Ricci tensor $Ric(\nero{\xi},\nero{\xi})$ are
\begin{equation}
\begin{split}
\tilde{E}_{1}&=\nero{\xi}=\left(
\begin{matrix}
\xi^{1}\\
\xi^{2}\\
\xi^{3}
\end{matrix}
\right)\\
\tilde{E}_{2}&=\left(
\begin{matrix}
\xi^{1}(q_{2}\xi^{2}+q_{3}\xi^{3})-q_{1}(\xi^{2}\xi^{2}+\xi^{3}\xi^{3})\\
\xi^{2}(q_{1}\xi^{1}+q_{3}\xi^{3})-q_{2}(\xi^{1}\xi^{1}+\xi^{3}\xi^{3})\\
\xi^{3}(q_{1}\xi^{1}+q_{2}\xi^{2})-q_{3}(\xi^{1}\xi^{1}+\xi^{2}\xi^{2})
\end{matrix}
\right)\\
\tilde{E}_{3}&=\left(
\begin{matrix}
q_{3}\xi^{2}-q_{2}\xi^{3}\\
q_{1}\xi^{3}-q_{3}\xi^{1}\\
q_{2}\xi^{1}-q_{1}\xi^{2}
\end{matrix}
\right)
\end{split}
\end{equation}
with the associated eigenvalues 
\begin{equation}
\begin{split}
\label{sectional_curvature_3dim}
\Lambda_{1}& =0\\
\Lambda_{2}& =\frac{E\kappa}{2 W^{3}}\\
\Lambda_{3}& =-\frac{\kappa}{4 W^{2}}\Big(\xi^{1}\xi^{1}(-4E+3\kappa q_{2}^{2}+3 \kappa q_{3}^{2})\\
&+\xi^{2}\xi^{2}(-4E+3\kappa q_{1}^{2}+3 \kappa q_{3}^{2})-6 \kappa q_{2}q_{3}\xi^{2}\xi^{3}\\
&+\xi^{3}\xi^{3}(-4E+3\kappa q_{1}^{2}+3 \kappa q_{2}^{2})-6 \kappa q_{1}\xi^{1}(q_{2}\xi^{2}+q_{3}\xi^{3})\Big) \ ,
\end{split}
\end{equation}
respectively.

With these eigenvectors the matrix for the basis transformation now is 
\begin{equation}
\nero{\Omega} = \left(
\begin{matrix}
\xi^{1} && \xi^{1}(q_{2}\xi^{2}+q_{3}\xi^{3})-q_{1}(\xi^{2}\xi^{2}+\xi^{3}\xi^{3}) &&q_{3}\xi^{2}-q_{2}\xi^{3}\\
\xi^{2} &&\xi^{2}(q_{1}\xi^{1}+q_{3}\xi^{3})-q_{2}(\xi^{1}\xi^{1}+\xi^{3}\xi^{3}) &&q_{1}\xi^{3}-q_{3}\xi^{1}\\
\xi^{3} && \xi^{3}(q_{1}\xi^{1}+q_{2}\xi^{2})-q_{3}(\xi^{1}\xi^{1}+\xi^{2}\xi^{2}) &&q_{2}\xi^{1}-q_{1}\xi^{2}
\end{matrix}
\right)  \ .
\end{equation}
In order to make the parallel transported reference frame orthonormal, we have to orthogonalise the above eigenvectors with respect to the Jacobi metric, thus for $i = 1, 2, 3$ we have
\begin{equation}
n_{i}^{2}\pieno{g}_{J}(\tilde{E}_{i},\tilde{E}_{i})=1
\end{equation}
namely
\begin{equation}
n_{i}=\frac{1}{\sqrt{2W \delta_{\alpha\beta} E_{i}^{\alpha}\,E_{i}^{\beta}}}
\end{equation}
These factors are
\begin{equation}
\begin{split}
n^{2}_{1}&=\frac{1}{2W(q)}\\
n^{2}_{2}&=\frac{1}{
	(q_{2}^{2}+q_{3}^{2})(\xi^{1})^{2
	}-2q_{1}q_{2}\xi^{1}\xi^{2}+(q_{1}^{2}+q_{3}^{2})(\xi^{2})^{2}-2q_{1}q_{3}\xi^{1}\xi^{3}
	-2q_{2}q_{3}\xi^{2}\xi^{3}+(q_{1}^{2}+q_{2}^{2})(\xi^{3})^{2}}\\
n_{3}^{2}&=\frac{(\xi^{1})^{2}+(\xi^{2})^{2}+(\xi^{3})^{2}}{((\xi^{1})^{2}+(\xi^{2})^{2})q_{3}^{2}-2q_{1}q_{3}\xi^{1}\xi^{3}-2q_{2}\xi^{2}(q_{1}\xi^{1}q_{3}\xi^{3})+q_{2}^{2}((\xi^{1})^{2}+(\xi^{3})^{2})+q_{1}^{2}((\xi^{2})^{2}+(\xi^{3})^{2})}
\end{split}
\end{equation}
The reference frame is composed by the normalized vectors 
\begin{equation}
E_{i}=n_{i}\tilde{E}_{i}
\end{equation}
and by representing  the Jacobi vector field as $J=J^{i}E_{i}$, we write the Jacobi-Levi Civita equations for the three components as
\begin{equation}\label{JLC Jacobi 3 dimensioni}
\begin{split}
&\frac{d^{2}J^{1}}{ds^{2}}=0\\
&\frac{d^{2}J^{2}}{ds^{2}}+\Lambda_{2}J^{2}=0\\
&\frac{d^{2}J^{3}}{ds^{2}}+\Lambda_{3}J^{3}=0
\end{split}
\end{equation}
Finally, passing to the physical time and by using Eq. \eqref{forma canonica} we get
\begin{equation*}
\ddot{f}^{k}+\omega_{(k)}(t)f^{k}(t)=0
\end{equation*}
with,  again, 
\begin{equation*}
\omega_{(k)}(t)=4W(q(t))^{2}\Lambda_{k}(t)-\frac{3}{4}\left(\frac{\dot{W}}{W}\right)^{2}+\frac{1}{2}\frac{\ddot{W}}{W} \ .
\end{equation*}
The results reported in the following are worked out by numerically integrating the following equations 
\begin{equation}\label{JLC Jacobi 3 dimensioni nel tempo}
\begin{split}
&\frac{d^{2}J^{1}}{dt^{2}}-\frac{\dot{W}}{W}\frac{dJ^{1}}{dt}=0\\
&\frac{d^{2}f^{2}}{dt^{2}}+\omega_{(2)}f^{2}=0\\
&\frac{d^{2}f^{3}}{dt^{2}}+\omega_{(3)}f^{3}=0
\end{split}
\end{equation}
with the following expressions for $\omega_{(2)}(t)$ and $\omega_{(3)}(t)$ \ :
\begin{equation}
\begin{split}
\omega_{(2)}(t)&:=\frac{2E \kappa}{W}-\frac{3}{4}\left(\frac{\dot{W}}{W}\right)^{2}+\frac{1}{2}\frac{\ddot{W}}{W}\\
\omega_{(3)}(t)&:=4W^{2}\Lambda_{3}-\frac{3}{4}\left(\frac{\dot{W}}{W}\right)^{2}+\frac{1}{2}\frac{\ddot{W}}{W}
\label{frequency_jacobimetric}
\end{split}
\end{equation}

\subsection{Numerical results}
\label{simulation3}
In Ref. \cite{cuervo2015non}, the alleged definitive argument to rule out the use of Jacobi metric to consistently describe the stability/instability of Hamiltonian dynamics was given by considering $N$ decoupled harmonic oscillators. The claim was that non-vanishing fluctuations of kinetic energy (due to the non affine parametrization of the arc length with time) entail parametric resonance in the JLC equation mimicking chaos for an integrable system. 
The authors considered the solutions of this system in the form 
\begin{equation}
q_{k}(t)= \cos\left(\omega t+\theta_k \right), \quad k=[1,N] \ ,
\label{qukappa}
\end{equation}
where $\theta_k = k\frac{2\pi f}{N}$, with the phases $\theta_k$ distributed on a fraction $f$ of the interval $2\pi$. 
It has been reported that the smaller $f$  the larger fluctuation of kinetic energy and the larger the Lyapunov exponent.
The fluctuation of kinetic energy $\sqrt{\sigma}$ is given in \cite{cuervo2015non} by
\begin{equation}
\label{flutuation_kineticenergy}
    \sqrt{\sigma}=\left(\frac{\sin(2 \pi f)}{\sqrt{2}N \sin(2 \pi f/N)}\right) \ .
\end{equation}
Notice that in the  $N\to\infty$, the kinetic energy fluctuation magnitude $\sqrt{\sigma}$ has a non-vanishing value, so that the authors claim that any dimension this basic integrable system would display non-physical instabilities.
In the next Section we will argue against this claim on the basis of an argument  related with the concentration of measure at high dimension.

 As shown in the preceding Section, and before in Refs.\cite{cerruti1996geometric,rick}, a consistent description of order and chaos is obtained using the Jacobi metric and writing the JLC equation for a parallel transported frame which is quite simple to be found for $N=2$, while the first non trivial extension is given by the $N=3$ case.
The JLC equation for the three dimensional case is given by Eqs.\eqref{JLC Jacobi 3 dimensioni nel tempo}. There are three principal directions of curvature, namely, the sectional curvatures [Eqs. \eqref{sectional_curvature_3dim}], identified by the planes generated by the velocity vector along a geodesic, $E_{1}=\nero{\xi}$, and the parallel transported  basis vectors $E_{k}$ with $k = 2, 3$. These sectional curvatures coincide with the eigenvalues of the operator $Ric(\nero{\xi},\nero{\xi})=R(\cdot,\nero{\xi})\nero{\xi}$; one of these is obviously zero because $Ric(\nero{\xi},\nero{\xi})\nero{\xi}=0$ while the other two are given by $\Lambda_{k}=\pieno{g}(E_{k},Ric(\nero{\xi},\nero{\xi})E_{k})$.

\begin{figure}[!htbp]
  \centering
  \begin{minipage}[t]{0.465\textwidth}
    \includegraphics[width=\textwidth]{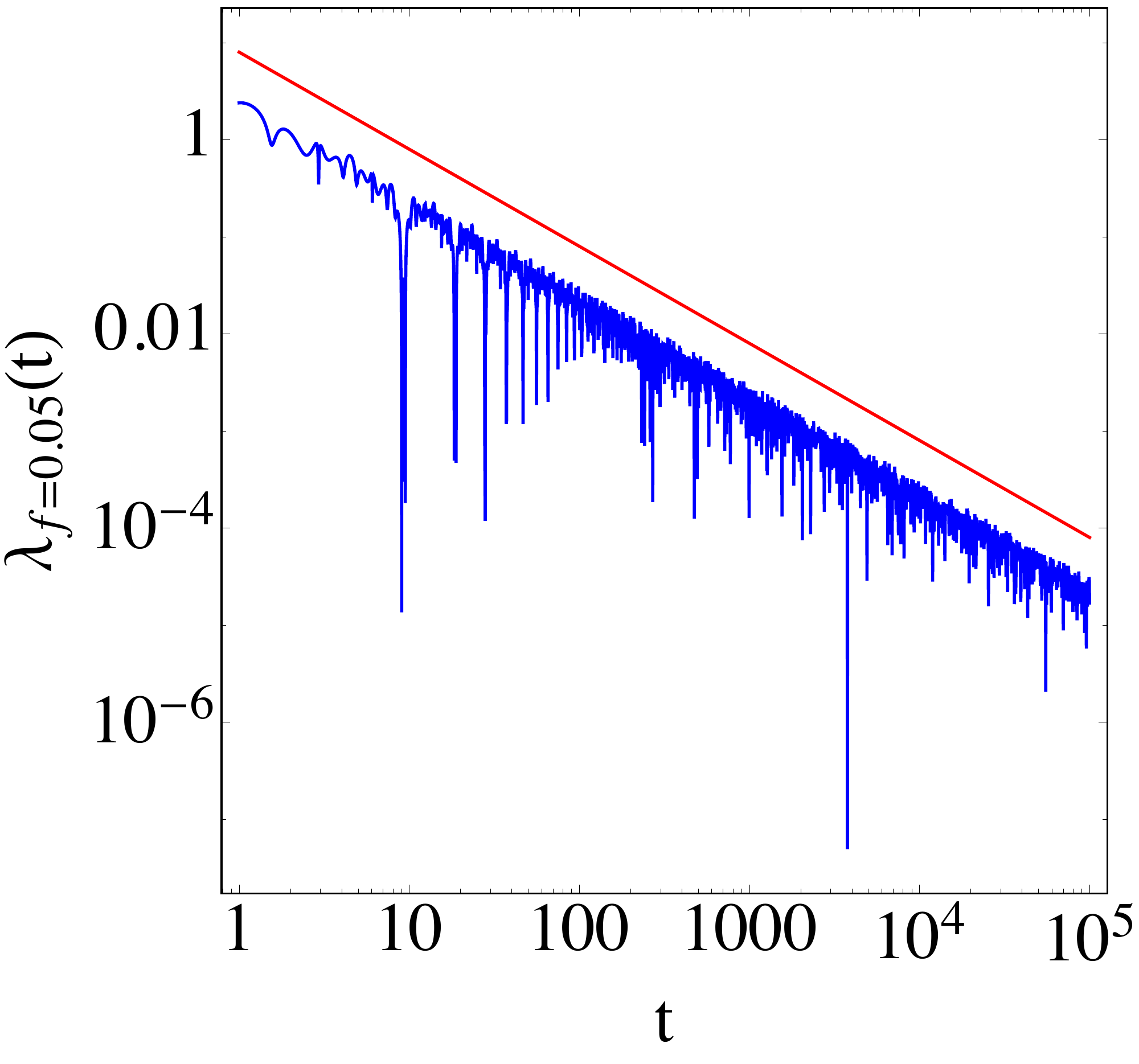}
    \caption{Comparison between the GLE (blue curve) and $1/t$ (red straight line). $f=0.05$, $\sqrt{\sigma(0.05)}=0.6968$.}
    \label{Lyap_Exp_nu0_05}
  \end{minipage}
  \hfill
  \begin{minipage}[t]{0.45\textwidth}
    \includegraphics[width=\textwidth]{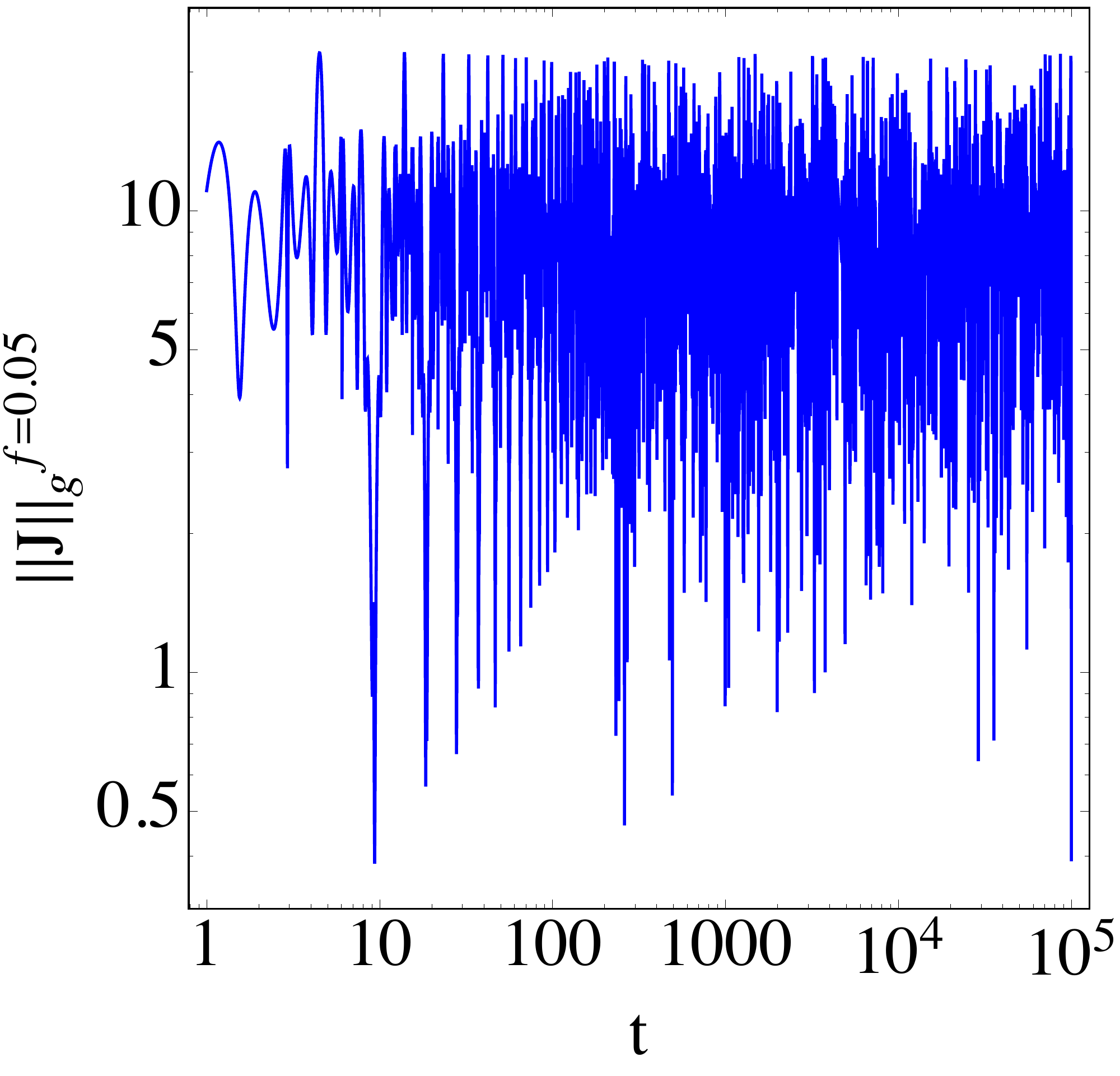}
    \caption{Norm of the Jacobi vector field. $f=0.05$, $\sqrt{\sigma(0.05)}=0.6968$.}
    \label{NormJ_nu0_05}
  \end{minipage}
\end{figure}

\begin{figure}[!htbp]
  \centering
  \begin{minipage}[t]{0.465\textwidth}
    \includegraphics[width=\textwidth]{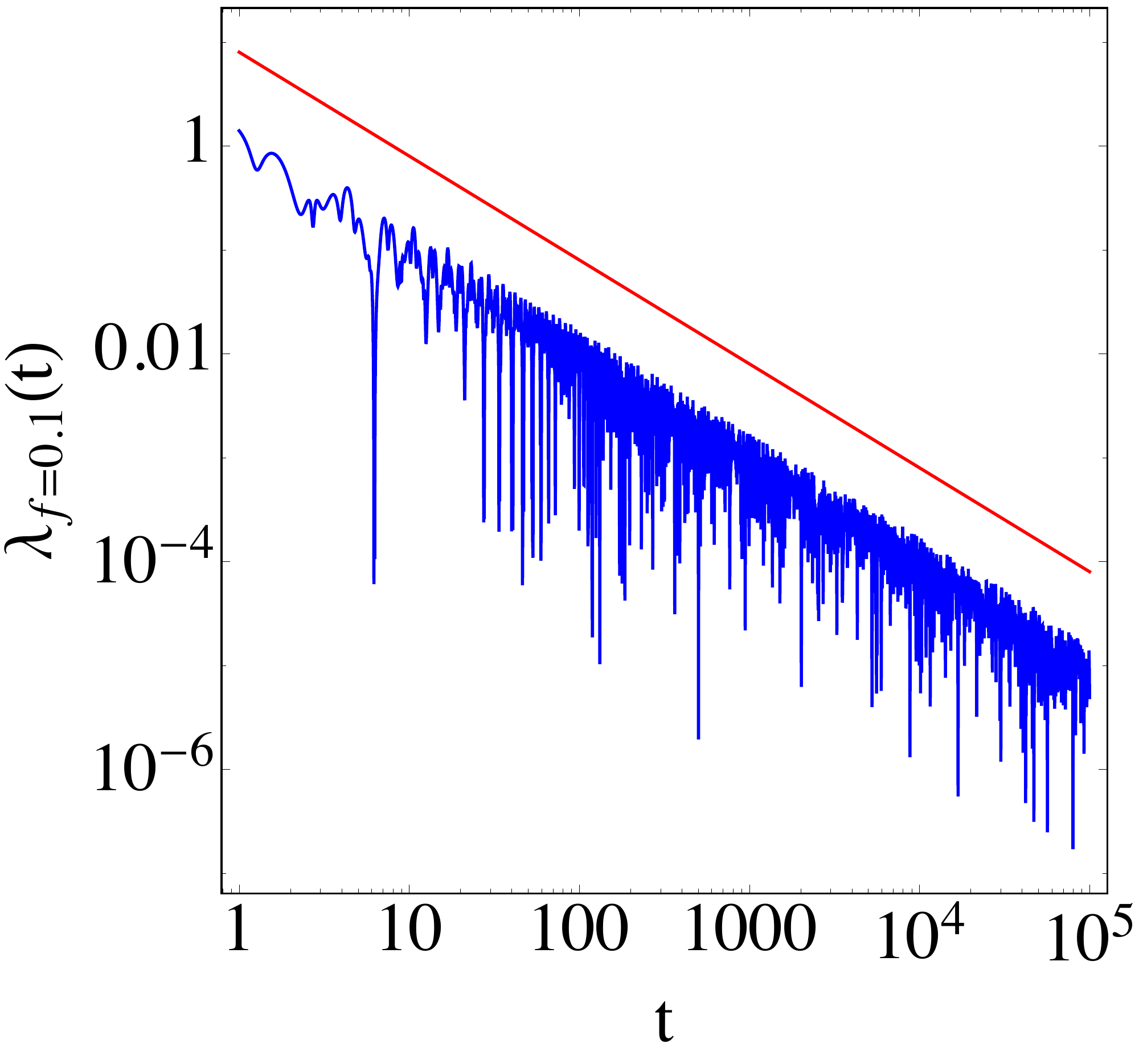}
    \caption{Comparison between the GLE (blue curve) and $1/t$ (red straight line). $f=0.1$, $\sqrt{\sigma(0.1)}=0.6663$.}
    \label{Lyap_Exp_nu0_1}
  \end{minipage}
  \hfill
  \begin{minipage}[t]{0.45\textwidth}
    \includegraphics[width=\textwidth]{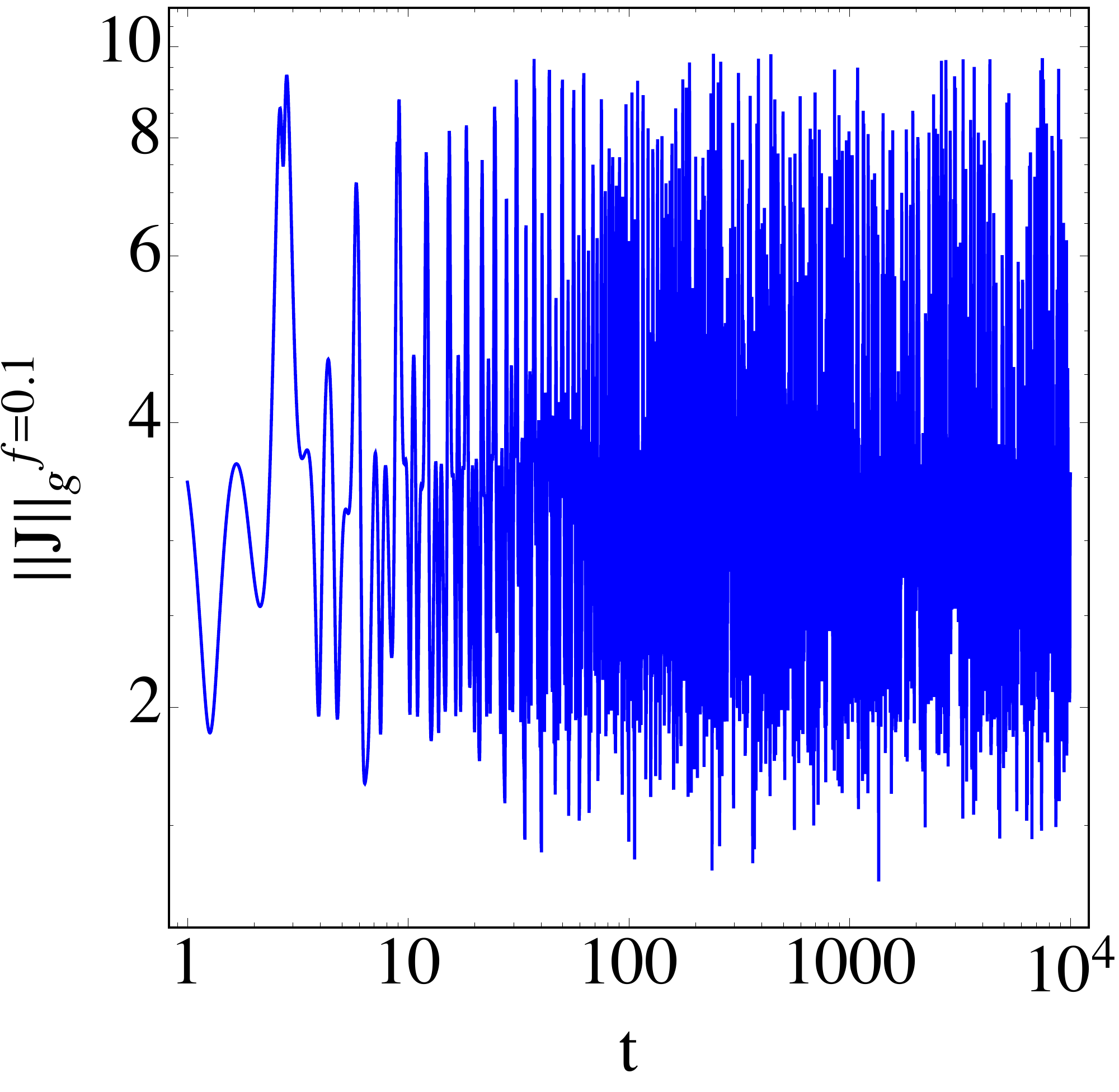}
    \caption{Norm of the Jacobi vector field. $f=0.1$, $\sqrt{\sigma(0.1)}=0.6663$.}
    \label{NormJ_nu0_1}
  \end{minipage}
\end{figure}

\begin{figure}[!htbp]
  \centering
  \begin{minipage}[t]{0.465\textwidth}
    \includegraphics[width=\textwidth]{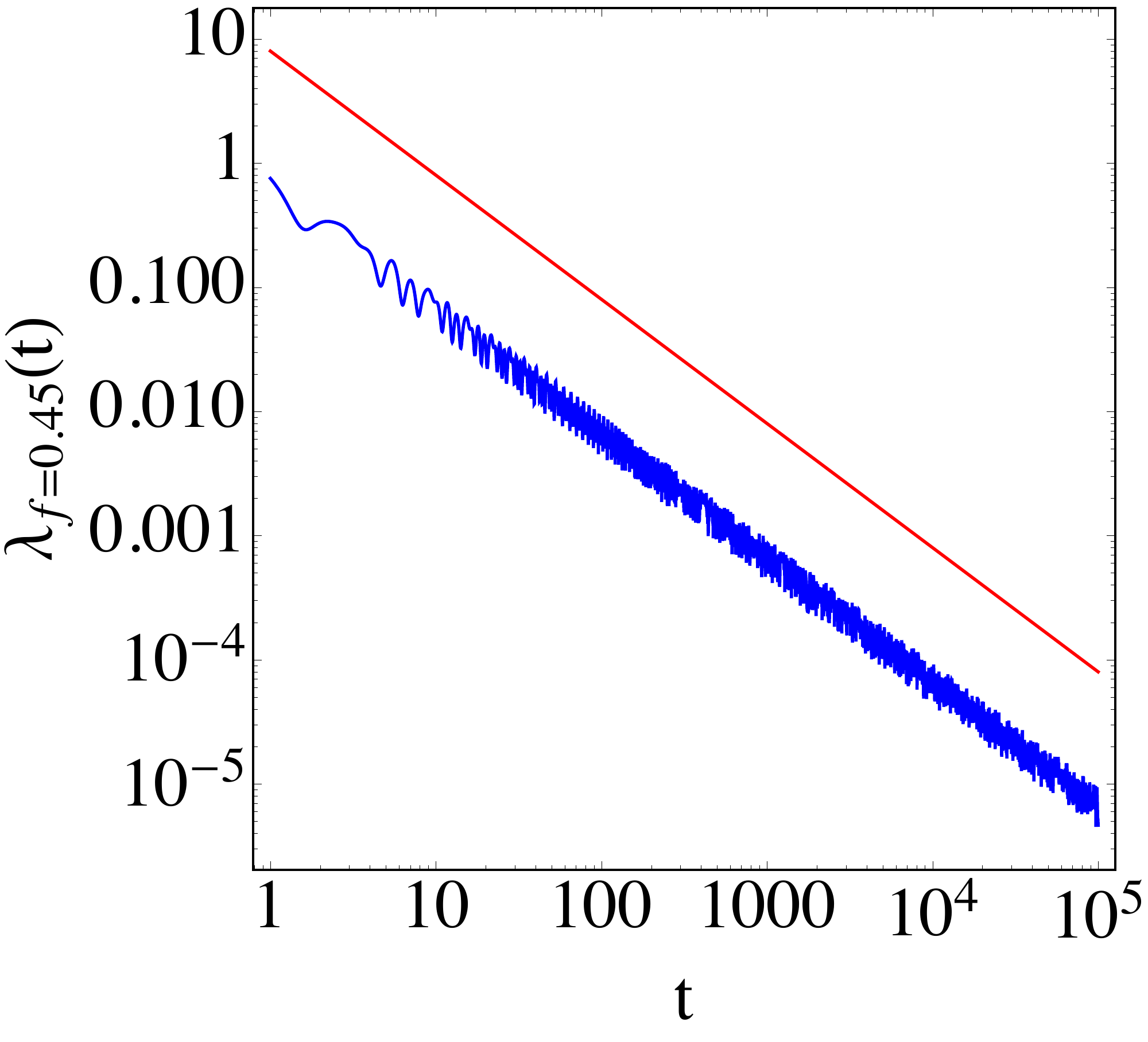}
    \caption{Comparison between the GLE (blue curve) and $1/t$ (red straight line). $f=0.45$, $\sqrt{\sigma(0.45)}=0.090$.}
    \label{Lyap_Exp_nu0_45}
  \end{minipage}
  \hfill
  \begin{minipage}[t]{0.45\textwidth}
    \includegraphics[width=\textwidth]{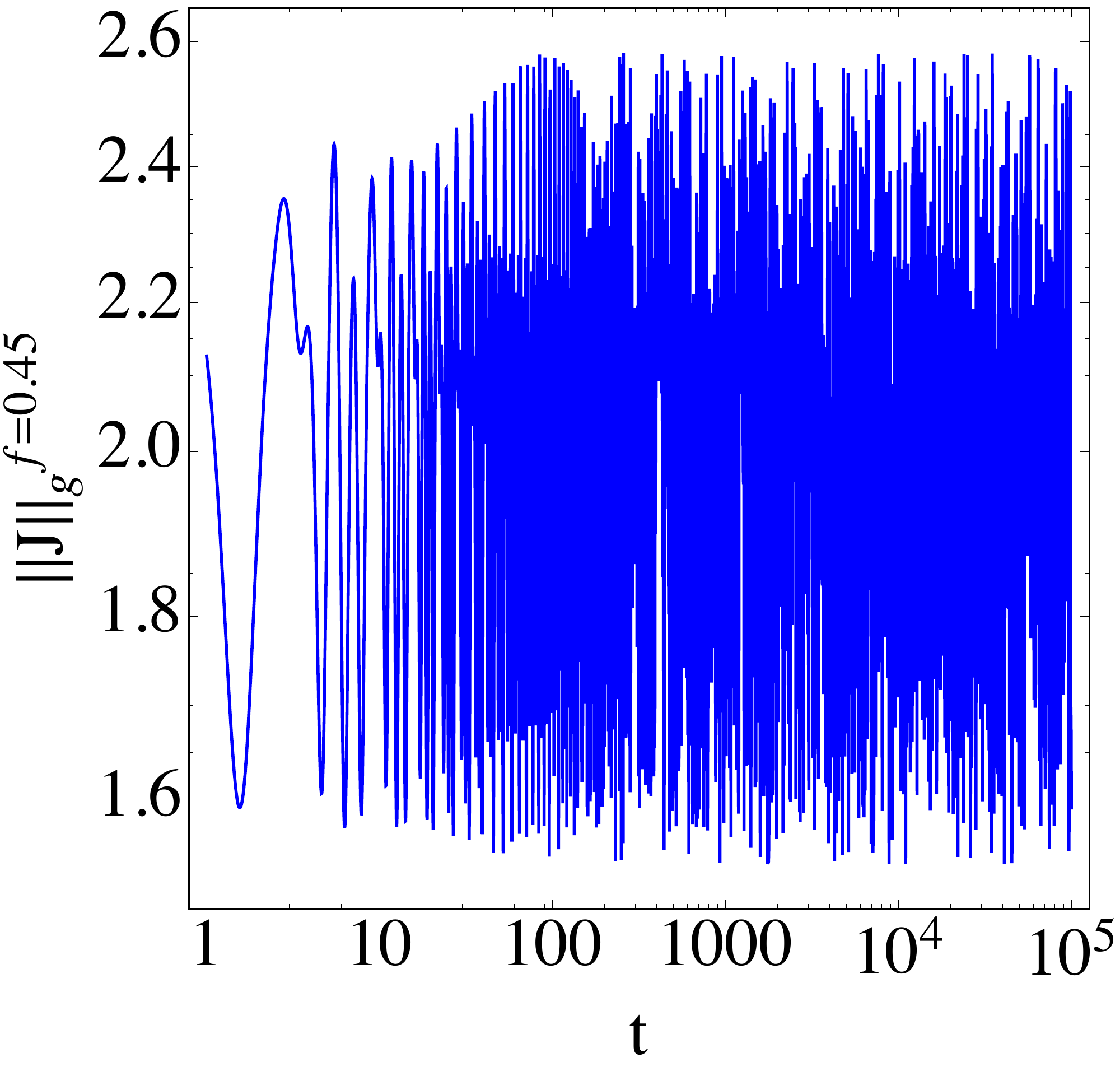}
    \caption{Norm of the Jacobi vector field. $f=0.45$, $\sqrt{\sigma(0.45)}=0.090$.}
    \label{NormJ_nu0_45}
  \end{minipage}
\end{figure}

In Figure 6 of Ref.\cite{cuervo2015non},  the largest value of $\lambda$ corresponds to a kinetic energy fluctuation level  $\sqrt{\sigma}\simeq 0.7$ which can be obtained with different values of $N$ and $f$.  
The Geometrical Lyapunov exponent and the norm of the Jacobi vector field have been worked out by numerically integrating Eqs.\eqref{JLC Jacobi 3 dimensioni nel tempo} for the following cases:  $N=3,\ f =0.05$ corresponding to $\sqrt{\sigma(0.05)}=0.6968$; for  $N=3,\ f =0.1$ corresponding to $\sqrt{\sigma(0.05)}=0.6663$; and for  $N=3,\ f =0.45$ corresponding to $\sqrt{\sigma(0.45)}=0.0900$. The outcomes are reported in Figures  \ref{Lyap_Exp_nu0_05} and \ref{NormJ_nu0_05}, in Figures \ref{Lyap_Exp_nu0_1} and \ref{NormJ_nu0_1}, and in Figures \ref{Lyap_Exp_nu0_45} and  \ref{NormJ_nu0_45}, respectively.

Equations \eqref{JLC Jacobi 3 dimensioni nel tempo} have been integrated with a fourth-order Runge-Kutta algorithm along the $q_k(t)$ given by \eqref{qukappa} and setting the phases $\theta_k$ uniformly distributed on a fraction $f$ of the interval $2\pi$.
It is well evident that  the norm of the Jacobi geodesic separation vector is always bounded, coherently with the decay with $1/t$ of the running value of $\lambda(t)$. The JLC equation written for the Jacobi metric  and with a parallel transported frame provides the correct result: no instability of the trajectories of an integrable system is found, contrary to the claim of Ref.\cite{cuervo2015non}.

%\newpage
\section{Concentration of measure of the volume occupied by accessible configurations with Jacobi metric}
\label{sec_concentration_measure}

We know from the work in Ref.\cite{cerruti1997lyapunov}  that the Jacobi-Levi Civita equation, written  in the natural reference frame for a large number of degrees of freedom, appropriately  works by producing Geometrical Lyapunov exponents in both qualitative and quantitative agreement with the standard Lyapunov exponents. 

In view of the above discussed problems due to the bouncing of phase trajectories/geodesics on the Hill's boundary, let us see why at large $N$ (large meaning just a few tens) 
the non physical divergencies are not found in spite of the use of the natural reference frame, in place of the parallel transported one, and in spite of the presence of kinetic energy fluctuations.

Consider a system composed by a large number of harmonic oscillators, denote by 
$\pieno{Q}=\{q_{1},\cdots,q_{N}\}$ and $\nero{P}=\{p_{1},\cdots,p_{N}\}$  the conjugate
momenta, with $\kappa =1$ for simplicity,  the Hamiltonian is 
\begin{equation}
H(\nero{P},\nero{Q})=\frac{\|\nero{P}\|^{2}_{{\Bbb R}^{N}}}{2}+\frac{\|\nero{Q}\|^{2}_{{\Bbb R}^{N}}}{2}=\sum_{i=1}^{N}\frac{1}{2}\left(|p_{i}|^{2}+|q_{i}|^{2}\right)
\end{equation}
and the Jacobi metric in the Hill's region $M_{E}$ is
\begin{equation}
\pieno{g}_{J}=2 \left[E-\frac{\|\nero{Q}\|^{2}_{{\Bbb R}^{N}}}{2}\right]\delta_{ij}\;dq^{i}\otimes dq^{j} \ .
\end{equation}
The associated Riemannian volume form is
\begin{equation}
\nu_{J}=\sqrt{det\,\pieno{g}_{J}}\;dq^{1}\wedge\cdots\wedge dq^{N}
\end{equation}
%The volume of configuration space $M_{E}:=\{q_{i}\in\Reale\,|\, E-V(\pieno{Q})>0\}$ is
%\begin{equation}
%\storto{V}_{N}:=\int_{M_{E}}d\mu=\int_{M_{E}}\sqrt{det\,\pieno{g}_{J}}\;dq^{1}\cdots dq^{N}
%\end{equation}
where the determinant of the metric is
\begin{equation}
det\,\pieno{g}_{J}=2^{N}\left(E-\frac{\|\nero{Q}\|^{2}_{{\Bbb R}^{N}}}{2}\right)^{N}\,.
\end{equation}
Therefore, the total volume of $M_{E}$ is given by the following integral
\begin{equation}
\label{eq:VolumeME_Jacobi}
\mathcal{V}_{N}(E)=2^{N/2}\int_{M_{E}}\left(E-\frac{\|\nero{Q}\|^{2}_{{\Bbb R}^{N}}}{2}\right)^{N/2}\;dq^{1}\cdots dq^{N}
\end{equation}
It is worth noticing a remarkable property of the Jacobi metric, its associated volume is proportional (up to an $N$-dependent factor) to the microcanonical ensemble measure
\begin{equation}
\label{eq:Gibbs_microcan_pf}
\Omega_{N,\mu}(E)=\int_{\mathcal{M}_{E}} |dp_{1}\wedge\ldots \wedge dp_{N}\wedge dq^{1}\wedge\ldots \wedge dq^{N}|
\end{equation}
where $\mathcal{M}_{E}=\left\{(\nero{P},\nero{Q})\in T^{*}M_{E}\,|H(\nero{P},\nero{Q})=E\right\}$. Due to
the quadratic form of the kinetic energy $2W=R^2$, the integral \eqref{eq:Gibbs_microcan_pf} can be rewritten as
\begin{equation}
\begin{split}
\Omega_{N,\mu}(E)&=\int_{M_{E}}\,\mathcal{A}(\mathbb{S}^{N-1})\,\,\int_{R\leq \sqrt{2(E-V(\nero{Q}))}}\,\,R^{(N-1)} dR\,\,dq^{1}\ldots dq^{N}=\\
&=\dfrac{\mathcal{A}(\mathbb{S}^{N-1})}{N} \int_{M_{E}}\,\,\left[2\left(E-V(\nero{Q})\right)\right]^{N/2}dq^{1}\ldots dq^{N}=\dfrac{\mathcal{A}(\mathbb{S}^{N-1})}{N}\,\,\mathcal{V}_{N}(E)\,\,\, ,
\end{split}
\end{equation}
where $\mathcal{A}(\mathbb{S}^{N-1})$ is the area of the unitary sphere.
Let us now consider  the volume of $M_E$ given by integral $\mathcal{V}_{N}(E)$  and show  that it \textit{concentrates} around
an $N-1$-dimensional manifold $\Sigma_{\bar{V}}=\left\{\nero{Q}\in M_{E}\,| V(\nero{Q})=\bar{V}\right\}$, in other words,  the overwhelming contribution to the volume integral  is given by microscopic configurations far from the boundary of $M_E$. Although it could be questionable to provide a statistical argument for an integrable
system for which ergodic hypothesis does not hold, the statistical averaging is intended as an averaging over all the possible configurations compatible with the constraint $H(\nero{P},\nero{Q})=E$.
It is convenient to rewrite
the integral \eqref{eq:VolumeME_Jacobi} with the volume element expressed in spherical coordinates 
\begin{equation}
|dq^{1}\wedge\cdots\wedge dq^{N}|=Q^{N-1}\sin^{N-2}(\phi_{1})\sin^{N-3}(\phi_{2})\cdots\sin(\phi_{N-2})d\phi_{1}\cdots d\phi_{N-2}\,dQ
\end{equation}
where $Q=\|\nero{Q}\|_{{\Bbb R}^{N}}=\sqrt{2 V(Q)}$, because integrating over the angular variables one obtains
\begin{equation}
\begin{split}
\mathcal{V}_{N}(E)&=2^{N/2} \mathcal{A}(\mathbb{S}^{N-1})\int_{0}^{\sqrt{2E}}\,\, \exp\left[\dfrac{N}{2}\ln\left(E-\dfrac{Q^2}{2}\right)+\left(N-1\right)\ln Q\right] dQ\\
&=2^{(N-1)/2} \mathcal{A}(\mathbb{S}^{N-1})\int_{0}^{E}\,\,dV \exp\left[N \left(\dfrac{1}{2}\ln\left(E-V\right)+\left(\dfrac{1}{2}-\dfrac{1}{N}\right)\ln V\right) \right]\\
&=2^{(N-1)/2} \mathcal{A}(\mathbb{S}^{N-1}) E^{N}\int_{0}^{1}\,\,dx \exp\left[-N F(x) \right]\,\,\,
\end{split}
\end{equation}
where $x=V/E$ is the relative value of the potential energy with respect to the total energy and 
\begin{equation}
F(x)=-\dfrac{1}{2}\ln\left(1-x\right)-\left(\dfrac{1}{2}-\dfrac{1}{N}\right)\ln x\,\,.
\end{equation}
As we are interested in the limit of large $N$, we can apply the Laplace approximation to evaluate the previous integral,
i.e. we consider the Taylor expansion around the minimum with respect to $x$ of $F$ in the interval $(0,1)$,
\begin{equation}
\int_{0}^{1} \exp\left[-N F(x) \right]\,\, dx\approx \exp\left[-N F(\bar{x})\right]\int_{0}^{1} \exp\left[-\dfrac{(x-\bar{x})^2}{2\sigma^2} \right] \,dx\,\,\,.
\end{equation}
where $\sigma^2=\left(F''(\bar{x}) N\right)^{-1}>0$. For a generic value of $N$, the solution of $F'(x)=0$ is
\begin{equation}
\bar{x}=\dfrac{1}{2}\dfrac{N-2}{N-1}\,\,\, ,
\end{equation}
which is actually a minimum since
\begin{equation}
\label{eq:d2fdx2_concmeas}
F''(\bar{x})=\dfrac{4(N-1)^3}{N^2(N-2)}>0 \qquad \text{for} \,\, N>2 \,\,.
\end{equation}
This means that the largest part of the volume $\mathcal{V}_N(E)$ is concentrated around the
hypersurface at constant potential energy $\Sigma_{\bar{x}E}$, a result very close to what is expected from the virial theorem \cite{note}.

From \eqref{eq:d2fdx2_concmeas} and $\sigma^2=\left(F''(\bar{x}) N\right)^{-1}$, it follows that the largest part of the volume ($\sim 99.7\% $) is concentrated around
$\Sigma_{\bar{x}E}$ in the interval $[\bar{x}-3\sigma,\bar{x}+3\sigma]$ with
\begin{equation}
\sigma=\dfrac{1}{2\sqrt{N}}\left[\dfrac{\left(1-\dfrac{2}{N}\right)}{\left(1-\dfrac{1}{N}\right)}\right]^{1/2}\approx \dfrac{1}{2\sqrt{N}}\,\,.
\end{equation}
This statistical argument shows that in the case of a large number of degrees of freedom 
the volume of the manifold is concentrated around a submanifold constant energy hypersurface $V=E/2$, far from the boundary
of the Hill region where Jacobi metric is singular. 

For example, with $N=100$ kinetic energy fluctuations of absolute value of $10\%$ occur with a probability of $68\%$ while the probability of configurations hitting the boundary ($E-V=0$) is $\sim 5.52\times 10^{-88}$.

\section{Discussion}
\label{conclusion}
Even though the point raised in Ref.\cite{cuervo2015non} is interesting, the conclusion  put forward by the authors is incorrect. The fluctuations of kinetic energy along a trajectory/geodesic of the Jacobi metric associated with an integrable system, like a collection of harmonic oscillators, are by no means responsible for the activation of parametric instability mimicking a chaotic behaviour. When the number of degrees of freedom of a Hamiltonian system is small, the associated geodesics can often approach the boundary $\partial M_E = \{ q\in M \vert V(q)=E\}$ of the mechanical manifold $M_E$, and, in so doing, the geodesics bounce on $\partial M_E$. The sharp reflection of the geodesics on the so-called Hill's boundaries \cite{montgomery2014s,giambo2014morse,giambo2015normal,seifert1948periodische} are at the origin of numerical instabilities which in principle could be perhaps avoided by a prohibitively high precision of the integration algorithm for the Jacobi--Levi-Civita equation describing the geodesic spread. However, throughout this paper we have shown that this problem can be fixed by choosing a parallel transported coordinate system. The stability/instability of geodesics is an intrinsic property thus in principle independent of the choice of the coordinate system, however, not all the coordinate systems are necessarily equivalent from the point of view of their numerical implementation and reliability of the corresponding outcomes. And, in fact, the sharp reflection of the geodesics by the boundaries $\partial M_E$ is accounted for by a sudden reflection of the coordinate axes of the parallel transported frames thus separating the true geometric origin of stability/instability of geodesics from the source of numerical artefacts related with their peculiar shape.    

We have then shown that when the number of degrees of freedom increases, then  the probability of approaching the boundary of the corresponding mechanical manifold $M_E$ gets lower and lower and, even if at finite $N$ the kinetic energy fluctuates it does not affect the strength of chaos measured through the outcomes of the JLC equation written for both the Jacobi and Einsenhart metrics which are in perfect agreement, as shown in Ref.\cite{cerruti1997lyapunov}. 
Thus already for a few tens of degrees of freedom the JLC equation for $(M_E,g_J)$ written in natural chart \cite{cerruti1997lyapunov,book}
\begin{eqnarray}
\frac{d^2J^k}{dt^2} &+& \frac{1}{E-V}\left(\partial_kV\delta_{ij}\frac{dq^i}{dt}
-  \partial_jV\frac{dq^k}{dt}\right)\frac{dJ^j}{dt} + [\partial^2_{kj}V]\ J^j
 \nonumber  \\
& + &  \frac{1}{E-V}\left[(\partial_kV)(\partial_jV)
-\left(\partial^2_{ij}
V+\frac{(\partial_iV)(\partial_jV)}{E-V}\right)\frac{dq^i}{dt}\frac{dq^k}{dt}
\right] J^j = 0~~.\nonumber
\label{jlc_final}
\end{eqnarray}
can be safely used, at most with the exclusion of a zero measure set of initial conditions. For very weakly coupled harmonic oscillators and $N=128$ these equations give $\lambda$ as small as $2\times 10^{-6}$ .

In conclusion, the study of order and chaos of Hamiltonian flows - identified as geodesic flows of the Jacobi metric in configuration space - is legitimate and coherent, although not unique.

%\newpage


\begin{references}
\bibitem{note}  The exact result expected from virial theorem,
	namely $\bar{x}=1/2$, is obtained considering the Boltzmann prescription for the microcanonical partition function
	\begin{equation}
	\Omega_{N,Bol}(E)=\int \delta(E-H(\nero{P},\nero{Q})) \,\, |dp_{1}\wedge\ldots\wedge dp_N\wedge q^{1}\wedge\ldots\wedge dq^{N}|\propto \int \left[2(E-V(\nero{Q})\right]^{\frac{N}{2}-1}\,\,| dq^{1}\wedge\ldots\wedge dq^{N}|\,\,.
	\nonumber \end{equation}
	This fact stands on the edge of a long standing debate about the "correct" prescription for the microcanonical partition function. Such a debate is out of the scope of the present work: we just report that Boltzmann prescription  gives a result more consistent with the virial theorem in the considered case.
\end{references}
\end{document}